\documentclass[twocolumn]{aastex631}
\usepackage{amsmath,amssymb}
\usepackage{booktabs}
\usepackage[figuresright]{rotating}
\usepackage{multirow,stackengine}
\usepackage{makecell}
\usepackage{capt-of}
\usepackage{ulem} 
\RequirePackage[normalem]{ulem} 
\RequirePackage{color}\definecolor{RED}{rgb}{1,0,0}\definecolor{BLUE}{rgb}{0,0,1} 
\newcommand{\massrate}{$M_{\odot}$\,yr$^{-1}$}

\newcommand{\hii}{H\textsc{ii}}
\newcommand{\msun}{$ M_\odot$}
\newcommand{\lsun}{$ L_\odot$}
\newcommand{\kms}{$\mathrm{km\,s}^{-1}$}

\newcommand{\degree}{$^{\circ}$}

\newcommand{\parcdeg}{\mbox{$.\!\!^{\circ}$}}

\newcommand{\hcnft}{HCN\,(4-3)}

\newcommand{\nthptt}{N$_2$H$^+$\,(3-2)}
\newcommand{\hcoptt}{HCO$^+$\,(3-2)}
\newcommand{\hcopoz}{HCO$^+$\,(1-0)}

\newcommand{\SE}{\texttt{SExtractor}}
\newcommand{\psk}{\texttt{PySpecKit}}
\newcommand{\astropy}{\texttt{Astropy}}
\newcommand{\montage}{\texttt{Montage}}

\received{March 3, 2023}
\revised{\today}
\accepted{June. 1, 2023}
\submitjournal{ApJS}

\shorttitle{HCN (4-3) Mapping of Massive Blue-profile Clumps} \shortauthors{Xu et al.}

\begin{document}

\title{Clump-scale Gas Infall in High-mass Star Formation: a Multi-transition View \\ with JCMT HCN (4--3) Mapping}

\correspondingauthor{Ke Wang (KW)}
\email{kwang.astro@pku.edu.cn}

\AuthorCollaborationLimit=10

\author[0000-0001-5950-1932]{Fengwei Xu}
\affiliation{Kavli Institute for Astronomy and Astrophysics, Peking University, Beijing 100871, People's Republic of China}
\affiliation{Department of Astronomy, School of Physics, Peking University, Beijing, 100871, People's Republic of China}

\author[0000-0002-7237-3856]{Ke Wang}
\affiliation{Kavli Institute for Astronomy and Astrophysics, Peking University, Beijing 100871, People's Republic of China}

\author[0000-0002-8760-8988]{Yuxin He}
\affiliation{Xinjiang Astronomical Observatory, Chinese Academy of Sciences, Urumqi 830011, People's Republic of China}
\affiliation{Key Laboratory of Radio Astronomy, Chinese Academy of Sciences, Urumqi 830011, People's Republic of China}

\author[0000-0001-7808-3756]{Jingwen Wu}
\affiliation{National Astronomical Observatories, Chinese Academy of Sciences, Beijing, 100101, People's Republic of China}
\affiliation{University of Chinese Academy of Sciences, Beijing, 100049, People's Republic of China}

\author{Lei Zhu}
\affiliation{CAS Key Laboratory of FAST, National Astronomical Observatories, Chinese Academy of Sciences, People's Republic of China}

\author[0000-0002-5065-9175]{Diego Mardones}
\affiliation{Departamento de Astronomía, Universidad de Chile, Casilla 36-D, Santiago, Chile}

\begin{abstract}


Gas infall motions play a crucial role in high-mass star formation and are characterized by observable signatures of blue-shifted asymmetric spectral line profiles (``blue profiles''). However, the connection between blue profiles and infall motions is unclear due to complex gas motions at parsec scales. In this study, we present the results of an \hcnft~mapping survey conducted with the JCMT, towards 38 massive clumps exhibiting blue profiles in \hcoptt. We extract 34 HCN cores from the 38 observed fields. The core-averaged spectra show various line profiles, indicating that blue-profile \hcoptt~does not guarantee the same in \hcnft. Through non-LTE radiation transfer calculations, we attribute the low detection rate of high-$J$ blue profiles to a combination of insufficient \hcnft~opacity and the intricate gas motion across different density layers. The comparison between the MALT90 and BGPS line surveys highlights the importance of appropriate tracers, high spectral resolution, and column density thresholds when searching for blue profiles. We select 11 reliable infall candidates and adopt the Hill5 model to fit the infall velocity of 0.2--1.6\,\kms, corresponding to 5\% to 74\% of free-fall velocity. Assuming spherically collapsing model, we estimate the median and mean mass infall rates to be $4.5\times10^{-3}$ and $7.6\times10^{-3}$ \massrate, respectively. The consistency of the mass infall rates among different transitions suggests a steady accretion process from the clump gas envelope to the inner region.


\end{abstract}


\section{Introduction}

Massive stars ($>8$\,\msun) play a predominant role in the energy budget of galaxies via their radiation, wind, and supernova events, but mass assembly processes including gas accretion or infall motions remain unclear. On the other hand, gravitational infall is a basic step in star formation theory \citep{Larson1969SF,Shu1987SF}, and expected in both ``core-fed'' \citep{McLaughlin1996TC,Mckee2003TC} and ``clump-fed'' massive star formation models \citep{Bonnell2001CA,Wang2010CA,VS2019GHC}, so identifying and studying the accretion flows which collect the material out of which stars form, either directly or indirectly, is an important aspect of understanding mass assembly of massive stars \citep{Fuller2005Infall,Sun2009Infall,Jackson2019MALT90}. 
Nevertheless, massive stars form in complex environments and large distances, thus features of individual cores embedded in the massive star forming clump are averaged together in the single-dish beam, making infall motions harder to observe \citep[e.g.][]{Reiter2011Infall,Liu2016EWBO,Yuan2017HMSC,Pillai2019,HuangBo2023} and observational evidence of collapse controversial to interpret \citep{Evans1991SF,Myers2000SF,Wu2003Infall,Wu2007Infall}. 

Self-absorbed, optically thick line profiles serve as phenomenological evidence of the infall within star-forming regions. When examining the emission arising from the infalling envelope positioned on the far side of the protostar, a proportional blue (Doppler) shift emerges, attributable to the velocity gradient toward the core. This blue-shifted emission evades absorption by foreground layers that are warmer or at a substantially different velocity \citep[see fig.1 in][]{Evans2003SF}, thereby leading to an excess of emission on the blueward side of the source velocity within the line profile \citep{Walker1986Infall,Walker1994Infall,Zhou1993Infall,Mardones1997Infall,Evans2003SF}. Notably, in instances where the source exhibits moderate optical thickness, a distinct blueward skew characterizes the line profile. Conversely, strongly self-absorbed sources manifest two discernible peaks, with the blue peak outshining the red peak to a moderate or significant degree. The depth of the self-absorption feature intensifies in the presence of substantial temperature gradients within the core, while the asymmetry of the line profile amplifies with pronounced velocity gradients \citep{Reiter2011Infall}. The distinctive line profile, commonly referred to as a ``blue asymmetric profile'' or simply ``blue profile'', enables the measurement and quantification of infall motion.

Various molecular species with different transitions, for example CS\,(2-1) by \citet{Sun2009Infall}, CS\,(3-2) by \citet{Zhang1998W51}, H$_2$CO\,(2-1) by \citet{Fuller2005Infall,Yoo2018Infall}, HCN\,(1-0) by \citet{Yang2020Infall}, HCN\,(3-2) by \citet{Wu2003Infall}, HCO$^+$\,(1-0) by \citet{Wu2007Infall,He2015Infall,He2016Infall,Jackson2019MALT90,Pillari2023HCOp}, HCO$^+$\,(3-2) by \citet{Reiter2011Infall}, HNC\,(1-0) by \citet{He2015Infall,He2016Infall,Saral2018MALT90}, THz NH$_3$ by \citet{Wyrowski2012Infall,Wyrowski2016Infall}, CH$_3$CN\,(19-18) by \citet{Liu2020Infall}, and CO\,(1-0) by \citet{Xu2021HGal} have been utilized to search for infall signatures in various environments in massive star-forming regions. In addition, comparisons of different tracers, including multiple transitions of the same tracer, have also been made through both observations \citep{Fuller2005Infall,Sun2009Infall,Yoo2018Infall,Xie2021HCOp} and simulations \citep{Chira2014Choice}, to explore which tracers are more efficient in revealing infall signatures in what kinds of sources. 

Above all, two major aspects can be improved in previous studies of infall motions in massive star formation: 1) since low-$J$ transitions can easily suffer from large optical depth and then be limited at low-density gas envelope \citep{Smith2012LineProfile}, \citet{Chira2014Choice} adopted radiative transfer calculations to show that high-$J$ transitions of HCO$^+$ and HCN offer the best combination of detectability of blue line profiles and visibility above typical noise levels, even better being \hcnft; 2) single-pointing observations cannot either rule out other possibilities which can also produce blue profiles \citep[e.g., outflow, rotation][]{Wu2007Infall} or resolve ``true'' core rather than clump-averaged collapse. Both aspects can be solved with the JCMT heterodyne array receiver program \citep[HARP;][]{Buckle2009HARP}. Designed to rapidly map extensive areas, HARP operates within the 325--375\,GHz frequency range and offers enhanced sensitivity in efficiently mapping \hcnft, providing an optimal strategy for investigating infall motions across a large sample.

Here, we present a JCMT HARP \hcnft~mapping survey of 38 massive clumps with known blue profiles in a pilot single-point \hcoptt~line survey conducted by \citet{Schlingman2011FollowLine,Shirley2013FollowLine}. The paper is organized as follows: Section\,\ref{sec:data} describes the sample selection, JCMT HARP observations and data reduction, and clump distance estimation. Results are presented in Section\,\ref{sec:result}. The discussions are followed in Section\,\ref{sec:discuss}. Finally, we give a summary and prospectus of the survey in Section\,\ref{sec:conclude}. 


\section{Data}
\label{sec:data}

\subsection{Sample Selection}
\label{data:sample}

\begin{figure}[!ht]
\centering
\includegraphics[width=1.0\linewidth]{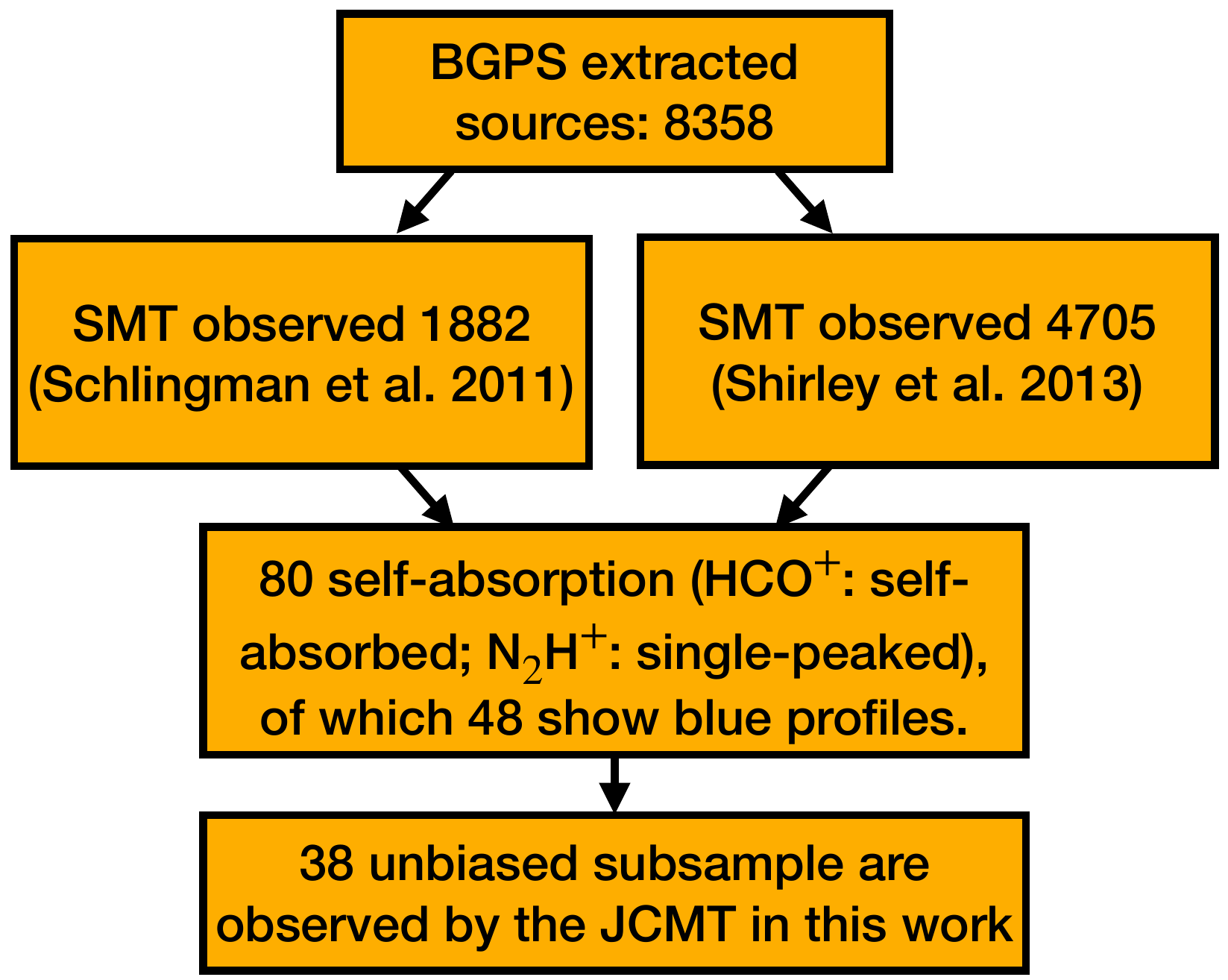}
\caption{The workflow of sample selection. Starting from the the 8358 BGPS sources, \citet{Schlingman2011FollowLine} and \citet{Shirley2013FollowLine} respectively performed line surveys, finally covering a total of 6194 BGPS sources. \citet{Shirley2013FollowLine} catalog 48 sources with blue profiles, of which 38 are observed by our JCMT \hcnft~mapping surveys. \label{fig:sampleflow}}
\end{figure}

The Bolocam Galactic Plane Survey (BGPS) imaged 170\,deg$^2$ sky 
at 1.1\,mm using Bolocam\citep[survey description in][]{Aguirre2011BGPSv1} and cataloged 8358 continuum clumps \citep[version 1.0.1 catalog;][]{Rosolowsky2010BGPSv1}. As a followup work, \citet{Schlingman2011FollowLine} and \citet{Shirley2013FollowLine} successively performed a single-pointed spectroscopy survey towards 1882 and 4705 BGPS clumps using the 10\,m Submillimeter Telescope (SMT) in \hcoptt~and \nthptt~with a spectral resolution of 1.1\,\kms.
\citet{Shirley2013FollowLine} then integrate and present a complete spectroscopic catalog of \hcoptt~and \nthptt~observations for 6194 sources in the BGPS v1.0.1 catalog between 7\parcdeg5$\le l \le$194\degree. 
Among the sample, 80 show self-absorbed line profiles where \hcoptt~shows two peaks and an absorption dip over the span of at least three channels (3.3\,\kms) with the \nthptt~line profile having a single-peak. Then, 48 are identified as blue asymmetric profiles, by comparing the optical thick \hcoptt~lines to the optically thin \nthptt~lines. These sources serve as excellent high-mass large-scale collapse candidates \citep{Shirley2013FollowLine}, which are the parent sample in our work. Due to the limit of observing time, a subsample of 38 clumps (including one adopted from the JCMT archive) are chosen as target fields (fields hereafter) in this work. The entire sample selection procedure is encapsulated in Figure\,\ref{fig:sampleflow}, elucidating the process through which the sample is curated, meticulously avoiding biases in relation to essential physical parameters such as distance, clump mass, or luminosity. It should be noted that the BGPS clumps provide an unbiased representation of the Galactic star-forming regions, affirming that the subsample maintains representativeness and consequently, the outcomes of this study hold a representative character.

All the fields are covered by legacy surveys of \textit{Spitzer}, \textit{Herschel}, and ATLASGAL, enabling us to obtain the infrared properties. We first retrieve the clump parameters including size, dust temperature, luminosity, mass and peak column density from \citet{Urquhart2018Property}, which are then corrected for by the updated distance (see Section\,\ref{data:distance}). The corrected clump-scale infrared properties are summarized in columns (8)--(12) of Table\,\ref{tab:sample}. The sample expands a wide range in: 1) evolutionary stages from infrared dark clouds (IRDCs) to infrared bright UC\hii~regions; 2) dust temperature from 9.7--34.4\,K; 3) mass from $1\times10^2$--$6\times10^3$\,\msun. 

\begin{figure*}[!ht]
\centering
\includegraphics[width=0.92\linewidth]{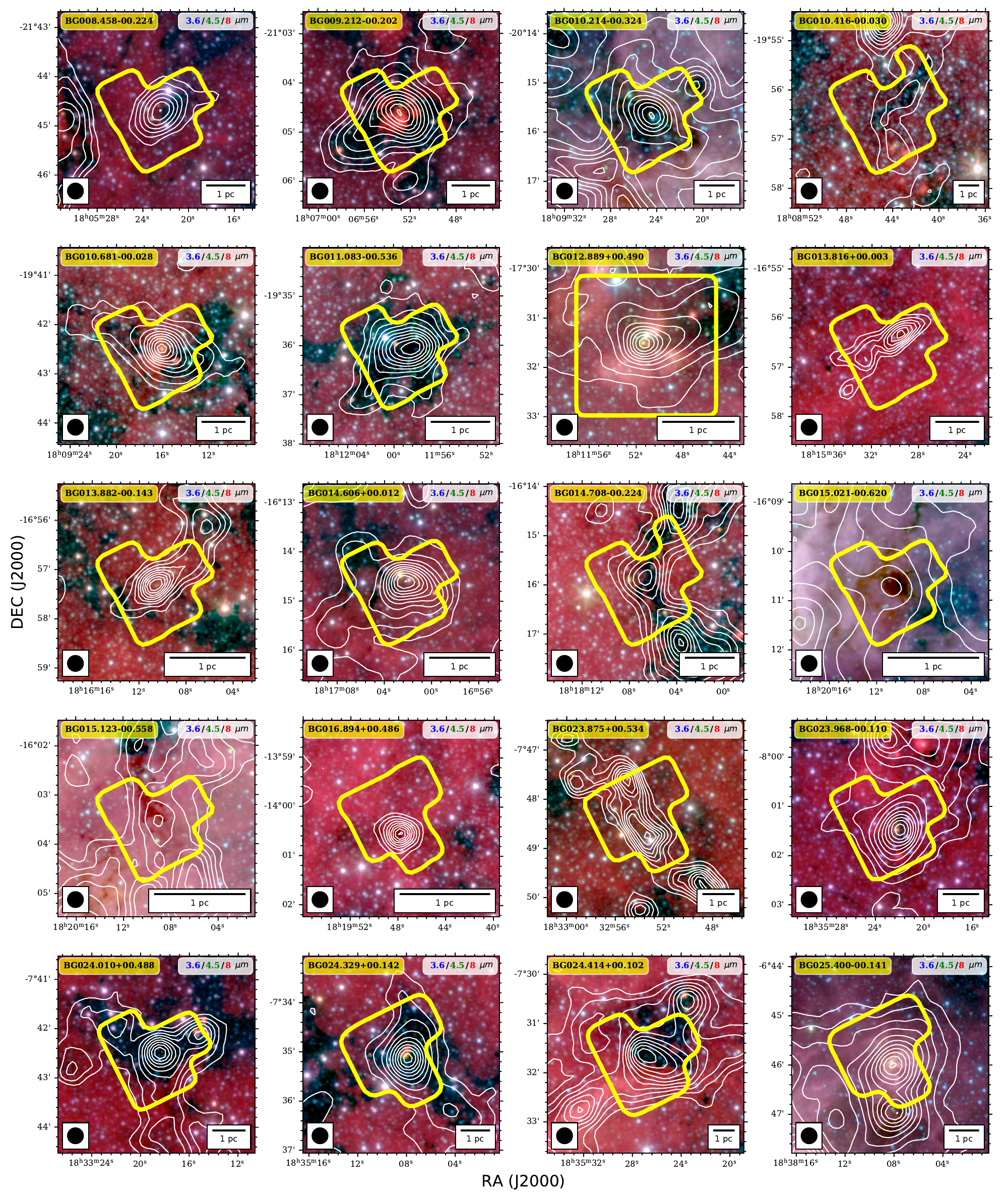}
\caption{Overview of the 38 clumps. Background shows the \textit{Spitzer} infrared three-color map (blue: 3.6\,$\mu$m; green: 4.5\,$\mu$m; red: 8\,$\mu$m). White contours are the ATLASGAL 850\,$\mu$m or the JCMT SCUBA-2 850\,$\mu$m continuum emission for three sources without ATLASGAL data. JCMT observing field is outlined by yellow line and the field name is shown at the upper left of each panel. The ATLASGAL/JCMT beam and the scale bar of 1\,pc are shown on the lower left and right, respectively. \label{fig:infrared}}
\end{figure*}
\addtocounter{figure}{-1}

\begin{figure*}
\centering
\includegraphics[width=0.92\linewidth]{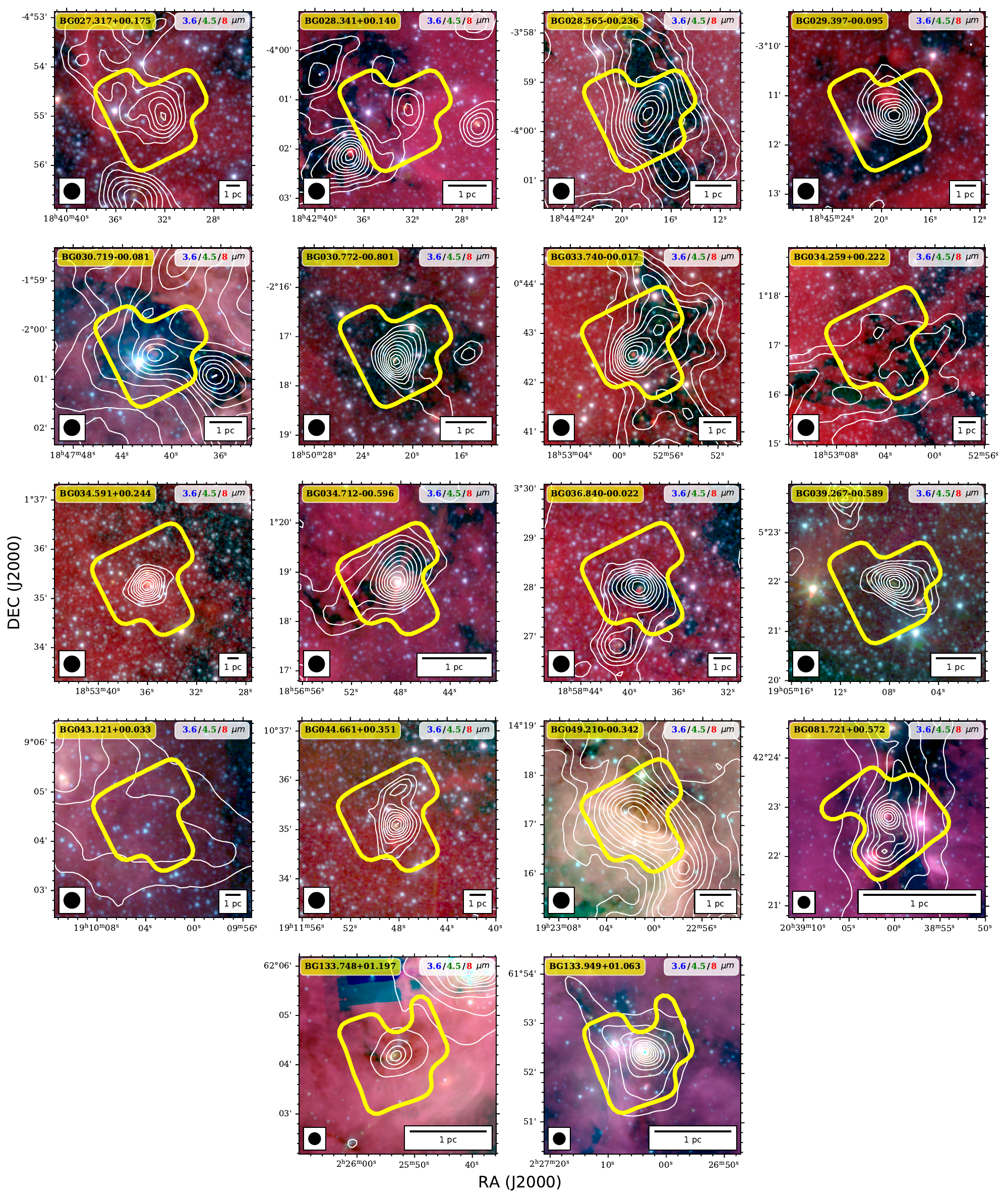}
\caption{Continued.}
\end{figure*}

\subsection{JCMT HARP Observations of HCN (4-3) and Data Reduction}
\label{data:obs}

The observations were carried out towards 38 blue-profile massive clumps with the 15\,m JCMT from 2019 October 13th to 2019 December 18th and from 2022 March 15th to 2022 June 5th (Project ID: M19BP033, M22AP051; PI: Ke Wang). The observation of BG012.889+00.490 (also IRAS 18089-1732) is retrieved from the JCMT archive (Project ID: M16AP067; PI: Hyunju Yoo).

We used the 16-pixel heterodyne array receiver program (HARP) for the front-end, and the Auto-Correlation Spectrometer and Imaging System (ACSIS) for the back-end \citep{Buckle2009HARP}. HARP is a single sideband receiver (SSB) comprised of a 16-receptor array arranged on a $4\times4$ grid. At the observing frequency, HARP has an angular resolution of 14\arcsec, and a main-beam efficiency of $\eta_{\rm mb}=0.61$. The footprint of the full array is $2\arcmin\times2\arcmin$. ``HARP5 Jiggle-Chop'' scanning mode is used to fill in the 30\arcsec~spacing between the receptors, therefore resulting in a $2\arcmin\times2\arcmin$ map with the pixel size of 6\arcsec, which is slightly over Nyquist sampling. The resultant scanning coverage for each field is highlighted by the yellow frame in Figure\,\ref{fig:infrared}. Note that two or three receptors are not operational in our observations, so the frames are usually incomplete squares except for BG012.889+00.490. ACSIS was set for a bandwidth of 250\,MHz with 8194 channels, centered at the frequency of \hcnft~after Doppler shift. A uniform channel width of $\sim0.03$\,MHz then leads to a velocity resolution of 0.026\,\kms. The position-switched mode was performed when the whole telescope moves away from the source and onto the reference position, which is specially chosen for each target based on the absence of CO and dust emission. During the observations, the weather condition had a precipitable water vapour (PWV) range of 1.575--2.575\,mm or a $\tau_{\rm 225\,GHz}$\footnote{The conversion from $\tau_{\rm 225\,GHz}$ to PWV is given in \citet{Dempsey2013SCUBA2}: $\tau_{\rm 225\,GHz}=0.04\times \mathrm{PWV}+0.017$.} of 0.08--0.12 (Band-3). The typical on-source time for each map is 40 minutes, or equivalently, 1.6 minutes for each HARP pixel.

The data were first calibrated and reduced by the pipeline introduced by \citet{Jenness2015JCMT}. The processed HARP-ACSIS data were converted into FITS format and then downloaded from the CADC's data collection\footnote{\href{https://www.cadc-ccda.hia-iha.nrc-cnrc.gc.ca/en/}{https://www.cadc-ccda.hia-iha.nrc-cnrc.gc.ca/en/}}. The orientations of the maps are determined by the K-mirror rotation which are different between observing fields, depending on the elevation of observation. To keep consistency, we regrid the maps to make y-axis aligned to the North. We convert the velocity in the barycentric frame to that in the local standard of rest (LSR). We smooth the velocity resolution to a uniform value of 0.2\,\kms~to enhance the signal-to-noise ratio (SNR) for further spectral line analyses. Achieved RMS noise level for each field is listed in column (9) of Table\,\ref{tab:sample}, with an averaged value of 0.10($\pm$0.02)\,K at a channel width of 0.2\,\kms.

\subsection{Distance Estimation}
\label{data:distance}

Reliable velocity determination is crucial for estimating a set of other physical properties of the clumps. We take advantage of the velocity at the local standard of rest ($V_\mathrm{LSR}$) derived from \nthptt~or \hcoptt~in \citet{Shirley2013FollowLine}.
We obey the following work flow to obtain the distance estimation for each source. First, we check for each source if any distance is already given in the references. If the distance is donated by kinematic distance or not given at all, we update the distance with the parallax-based Bayesian maximum-likelihood distance estimation approach version 2.4.1 \citep{Reid2016KD,Reid2019KD}. Note that if one source is located outside the solar circle (i.e., the distance from the Galactic Center $R_{\rm gc}>8.5$\,kpc) or is at a tangential point, we will calculate one unique distance. However, if one source is located within the solar circle (i.e. $R_{\rm gc}<8.5$\,kpc), two possible distances are obtained (one near, one far). This degeneracy is commonly referred to as kinematic distance ambiguity (KDA). To address KDA, we follow the methods described in \citet{Urquhart2018Property}, which test several criteria one by one to determine the distance. First, we search the SIMBAD database for any previous distance estimation. We then choose the one closest to the value reported in the literature. If no reference is found, then we check whether the source elevation ($z$) to the Galactic mid-plane\footnote{The Sun is $10\pm2$\,pc higher than the Galactic mid-plane \citep{Griv2021Sun}, therefore a southward shift is included.} for the farther distance is larger than 120\,pc. If this is the case, then the closer one is adopted. After the above workflow, the distances and their references are listed in columns 5--6 of Table\,\ref{tab:sample}. 


\section{Results} \label{sec:result}

\subsection{Detection of HCN (4-3) Emission} \label{result:detection}

We generate the moment zero (M0) maps to show how the emission of \hcnft~is distributed. For each field, we first extract the velocity range of [$V_{\rm LSR}-\mathrm{d}V, V_{\rm LSR}+\mathrm{d}V$] where $\mathrm{d}V=10$\,\kms~to cover the majority of HCN line emission. Then we integrate the spectra within the velocity range at each pixel and obtain the M0 maps shown in Figure\,\ref{fig:m0-1/2}. 

Most of the HCN emission shows core-like condensed structures, although some show more extended and irregular ones. We then adopt an automatic source extraction algorithm \SE\footnote{\href{https://sextractor.readthedocs.io/en/latest/Introduction.html}{https://sextractor.readthedocs.io/en/latest/Introduction.html}.} \citep{1996A&AS..117..393B} on M0 maps to extract HCN emission sources. The advantages of \SE~in our case are: 1) to reduce background emission; 2) to support local rms noise input to serve as pixelwise thresholds; 3) to deblend the potentially blended sources in one field. The algorithm procedure and the parameter settings are described in Appendix\,\ref{app:sextractor} in detail. As a result, a total of 34 HCN sources are extracted and fitted by 2D Gaussian profiles shown by green solid ellipses in Figure\,\ref{fig:m0-1/2}. Since HCN sources have physical sizes of 0.08--0.35\,pc (column 9 of Table\,\ref{tab:HCNcores}) which are much smaller than the massive clumps of $\sim1$\,pc, we define them as HCN ``cores'' hereafter. For further spectral line analyses, we also assigned a circle with diameter of 30\arcsec~\citep[which is the beam size of SMT at 270\,GHz][]{Shirley2013FollowLine} to the six fields where no HCN was detected, shown as green dashed circles in Figure\,\ref{fig:m0-1/2}. The basic fitted parameters of the 34 HCN cores including offsets (along x and y axes) from the field center, major and minor axes ($\theta_{\rm maj}$ and $\theta_{\rm min}$), position angle (PA), and peak flux ($F_\mathrm{peak}$) are listed in Column 3--8 of Table\,\ref{tab:HCNcores}. 

Following the method of \citet{Rosolowsky2010BGPSv1} and \citet{Contreras2013CSC}, the deconvolved angular radius is written as,
\begin{equation}\label{eq:theta_source}
    \theta_{\rm core} = \eta\left[\left(\sigma^2_{\rm maj}-\sigma^2_{\rm bm}\right)\left(\sigma^2_{\rm min}-\sigma^2_{\rm bm}\right)\right]^{1/4},
\end{equation}
where $\sigma_{\rm maj}$ and $\sigma_{\rm min}$ are calculated from $\theta_{\rm maj}/\sqrt{8\ln2}$ and $\theta_{\rm min}/\sqrt{8\ln2}$, respectively. The $\sigma_{\rm bm}$ is the averaged dispersion size of the beam (i.e., $\theta_{\rm bmaj}/\sqrt{8\ln2}$, where $\theta_{\rm bmaj}\simeq14\arcsec$ is the JCMT beam at the frequency of \hcnft). $\eta$ is a factor that relates the dispersion size of the emission distribution to the determined angular radius of the object. We have elected to use a value of $\eta=2.4$, which is the median value derived for a range of models consisting of a spherical, emissivity distribution \citep{Rosolowsky2010BGPSv1}. Therefore, the physical size of the core is derived from $R_{\rm core} = \theta_{\rm core} \times D$ ($D$ is the distance), which is listed in column 9 of Table\,\ref{tab:HCNcores}. Some of the cores have sizes comparable to the beam size, rendering them unresolved. In these cases, Column 9 of the corresponding rows is marked with ``--'' as a notation.

\subsection{Averaged Spectra from HCN Cores}
\label{result:avgspectra}

\begin{figure*}[!ht]
\centering
\includegraphics[angle=0,width=1.0\linewidth]{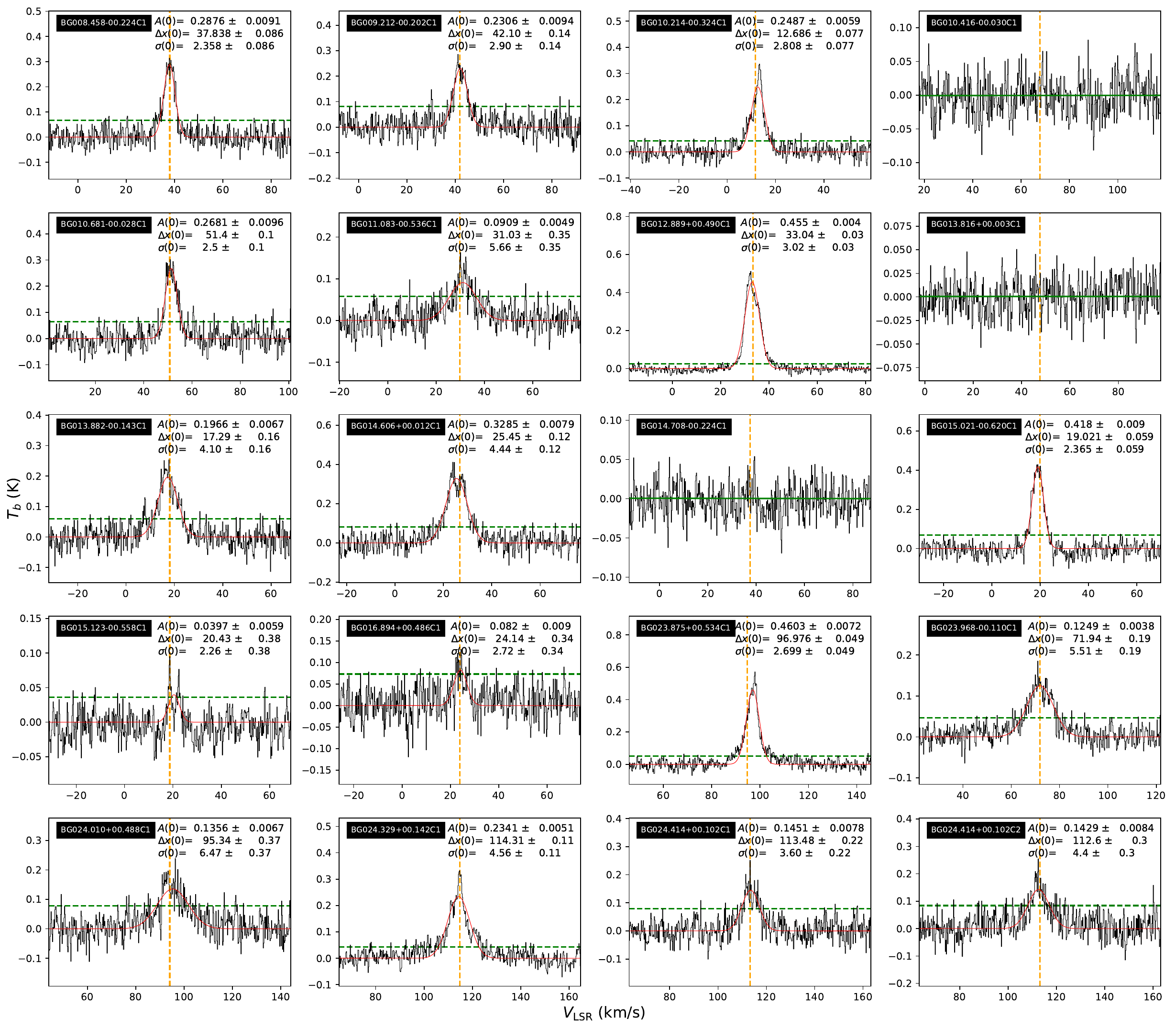}
\caption{Averaged \hcnft~lines from the defined HCN cores whose name are labeled on the top left of each panel. For solid detections, the \hcnft~lines are fitted by Gaussian profiles. The $2\sigma$ threshold of the best-fitting model is shown with a green dashed line and a red line, respectively. The results of the Gaussian fitting (amplitude $A$, centroid velocity $\Delta x$, and velocity dispersion $\sigma$) are shown on the top right. Non-detection spectra are not fitted, and only the baselines (green horizontal lines) are shown. The systematic velocities in previous surveys are marked with orange dashed lines. \label{fig:Gaussian_Fitter}}
\end{figure*}
\addtocounter{figure}{-1}

\begin{figure*}[!ht]
\centering
\includegraphics[angle=0,width=1.0\linewidth]{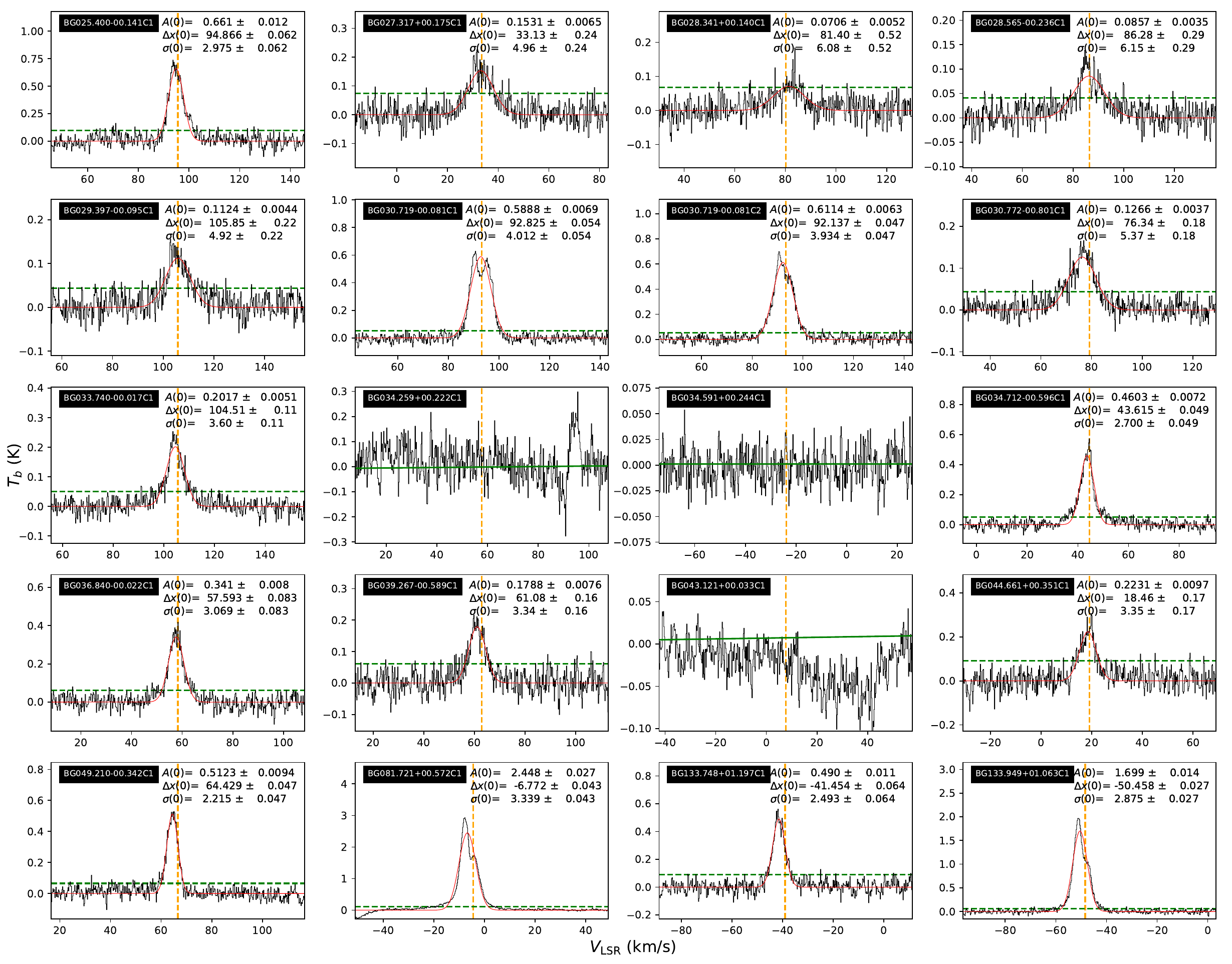}
\caption{Continued.}
\end{figure*}

The \hcnft~lines are extracted from the defined regions (including 34 HCN cores and 6 circles defined in Section\,\ref{result:detection}). We first smooth the velocity resolution to a uniform value of 0.2\,\kms, to enhance the signal-to-noise ratio (SNR) for further spectral line analyses. Then the baseline of spectra are subtracted and the baseline-free spectra are shown in Figure\,\ref{fig:Gaussian_Fitter}. The SNRs are defined as the ratio of $T_{\rm peak}$ to $\sigma$. In our analyses, the spectra with low SNR ($<2$) are classified as non-detections of HCN emission, while others are solid detections. We also visually double-check the spectra to exclude the potential temperature jump at bad channels. We note that although the field BG034.259+00.222 has a detection in the North, the \hcnft~line has a large velocity deviation ($\sim40$\,\kms) from the systematic velocity. In addition, the detected core BG034.259+00.222C1 is near the edge of the field. So, we assume that BG034.259+00.222C1 is not correlated with the clump and excluded in the further discussion. Another notice is that although BG015.123-00.558 has no detection in the field, the averaged spectrum from the central circle shows a SNR$\sim2$ detection of emission. The non-detection in the \SE~algorithm should be due to extended and diluted emission. As a result, 34 of 40 spectra show solid HCN detection and six are designated as non-detection spectra.

For the HCN spectra with solid detection, we fit them with a single Gaussian model by the Python package \psk, which are shown on the upper right corner of each panel in Figure\,\ref{fig:Gaussian_Fitter}. The Gaussian parameters, including amplitude, centroid velocity, and linewidth, as well as their uncertainties are listed in columns (2)--(4) of Table\,\ref{tab:HCNpar}. For the six non-detection spectra, columns (2)--(4) of Table\,\ref{tab:HCNpar} are filled with ``-''. We also flag the non-detection spectra with ``N'' in column (9).

\subsection{Synergy with Previous Line Surveys} \label{result:synergy}

Line surveys conducted by \citet{Schlingman2011FollowLine,Shirley2013FollowLine} not only serve as a guide for our follow-up survey (see Section\,\ref{data:sample}), but also provide a large legacy value for spectral analyses in our work. The \nthptt~lines are observed to be optically thin \citep{Shirley2015Lines}, which can therefore be used to determined the systematic velocity and velocity dispersion of massive clumps.

Two important caveats warrant consideration in our analysis. Firstly, the \nthptt~lines were observed using the SMT, whose beam size is approximately twice that of the JCMT. Consequently, the \nthptt~lines may reflect the systematic velocity of the entire clump or the dense inner region within the clump, rather than the velocity of the central dense core as indicated by the \hcnft~lines. In essence, the coherence in velocity between parent clumps and HCN cores should underpin our discussions concerning line profiles, as discussed in Section\,\ref{result:profiles}.

Secondly, due to the larger physical coverage by the SMT beam, encompassing approximately four times the area of the JCMT beam, more turbulent motion should be included. Consequently, broader line widths are anticipated. This implies that the \nthptt~lines should be narrower than they would be if observed within the JCMT beam. When comparing with the JCMT results in Section\,\ref{result:profiles}, we should always keep in mind that the line width of the \nthptt~line could be overestimated.

\begin{figure}[!ht]
\centering
\includegraphics[width=1.0\linewidth]{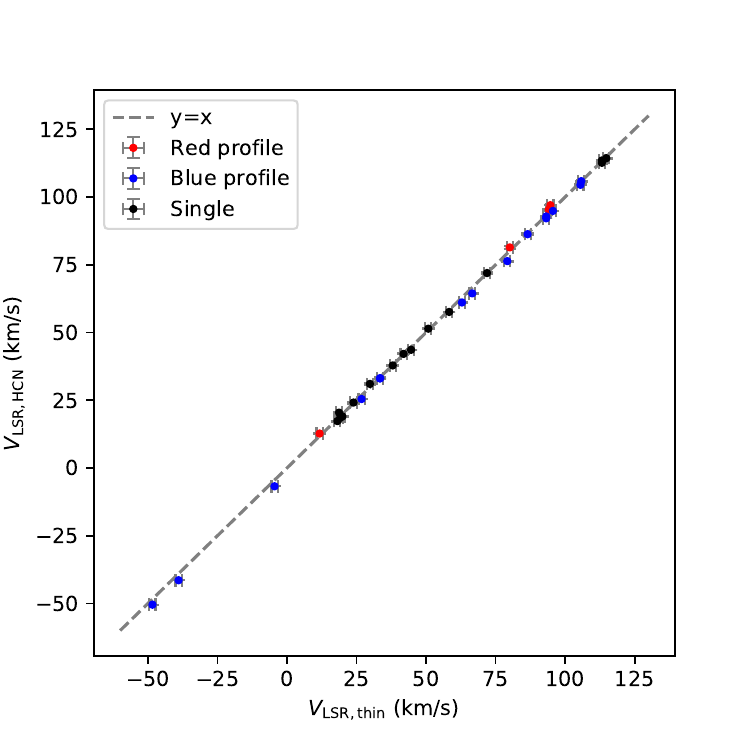}
\caption{Consistency between $V_{\rm LSR, HCN}$ and $V_{\rm LSR, thin}$. Red: red-profile spectral lines; blue: blue-profile spectral lines; black: no clear line profiles. The errorbars at two directions are given by spectral line fitting errors. \label{fig:consistency_vlsr}}
\end{figure}

Here, we check the consistency between the velocity derived from \hcnft~lines $V_{\rm LSR,HCN}$ and that fitted by optical thin lines \nthptt~$V_{\rm LSR,thin}$ from \citet{Shirley2013FollowLine}. As shown in Figure\,\ref{fig:consistency_vlsr}, $V_{\rm LSR,HCN}$ and $V_{\rm LSR,thin}$ always share the same value within the uncertainty, indicating a good correspondence between two surveys, which establishes the basis for the blue profile analyses in this paper.

\subsection{Variety of Observed Spectral Line Profiles}
\label{result:profiles}
As shown in Figure\,\ref{fig:Gaussian_Fitter}, the averaged \hcnft~line shapes differ from core to core, some showing asymmetric profiles or double-peak profiles (non-Gaussian). To distinguish line profiles and study the distribution statistically, we adopt the definition of velocity difference by \citet{Mardones1997Infall},
\begin{equation}\label{eq:velocity_difference}
    \delta V = \frac{V_\mathrm{HCN,peak}-V_\mathrm{sys}}{\mathrm{d}V_\mathrm{thin}},
\end{equation}
where the difference between the peak velocity of \hcnft~$V_\mathrm{HCN,peak}$ and systematic velocity derived from optically thin line $V_{\rm sys}$ are normalized by the FWHM of the thin line $\mathrm{d}V_{\rm thin}$. The normalization makes it convenient and robust to set a uniform criterion to distinguish different line profiles, especially for sample with a wide range of line widths. 

We first calculate the velocity at the peak intensity as $V_\mathrm{HCN,peak}$, which is listed in column (5) of Table\,\ref{tab:HCNpar}. To obtain $V_{\rm sys}$, we then retrieve the fitting results of \nthptt~lines from \citet{Shirley2013FollowLine} where the \nthptt~lines are thought to be optically thin and taken as tracers of systematic velocity. $V_{\rm sys}$ for each core is marked as an orange dashed line in Figure\,\ref{fig:Gaussian_Fitter}. If two cores are in one clump, then they share the same $V_{\rm sys}$ and $\mathrm{d}V_{\rm thin}$, which are listed in columns (6)--(7) of Table\,\ref{tab:HCNpar}. By Eq.\,\ref{eq:velocity_difference}, normalized $\delta V$ is then calculated and listed in column (8). We designate those with $\delta V<-0.25$ as significant blue profile (donated with ``BP'' hereafter), those with $\delta V>0.25$ as significant red profile (``RP'' hereafter), and those with $-0.25<\delta V<0.25$ as single component (``S'' hereafter) that have insignificant asymmetric profiles. The designation is listed in column (9) of Table\,\ref{tab:HCNpar}. 

We note that the second caveat in Section\,\ref{result:synergy} can cause underestimation of $\delta V$ due to systematic overestimation of $\mathrm{d} V_{\rm thin}$ as described in Eq.\,\ref{eq:velocity_difference}. Consequently, there exists the possibility of bias, where the criteria for defining red or blue profiles (i.e., $\delta V>0.25$ or $<-0.25$) might be more stringent than intended. This could potentially classify marginally satisfactory line profiles as non-asymmetric, resulting in a bias that reinforces the definition of pronounced line profiles but may also elevate the false negative rate for weak line profiles. To address this potential bias, a secondary assessment should be conducted through visual inspection. Two instances, BG009.212-00.202C1 and BG023.968-00.110C1, exhibit blue-shifted double peaks with $\delta V$ values of $-0.13$ and $-0.14$ respectively. Despite not meeting the $\delta V$ threshold, they are designated as ``BP'' due to their distinctive characteristics. Additionally, BG027.317+00.175C1, while satisfying the blue-profile criterion, possesses a low signal-to-noise ratio, which is then labeled as ``S''. Besides, BG030.772-00.801C1 and BG049.210-00.342C1, despite having $\delta V<-0.25$, each features only a single peak. As a result, they are then classified as ``S''. Consequently, the final identification designates 14 cores as ``BP'' (referred to as HCN-BP cores) and four cores as ``RP''.

\subsection{Infall Candidates Identified by Line Mapping}
\label{result:candidate}

Statistically, infall motion is the most likely interpretation of the observed blue profiles. However, in individual cases, it is not the only possibility. Rotation and outflows can also produce blue profiles \citep[e.g.,][]{Wu2003Infall,Wu2007Infall}. Resolved mapping observations are needed to investigate the nature of blue profiles.
Rotation of core always exhibits blue and red profiles at different spatial positions, which are mistaken blue profiles at single-pointed observation. In a similar way, outflow lobes are easily ruled out if red-shifted emission is predominately from an extended wing. A profile that survives these tests provides a strong indication of infall, and the source can be seen as a candidate for collapse. 

To provide with a better visualization of mapping observations, we present spectral line grids for each HCN-BP cores in Figure\,\ref{fig:specgrid} to exclude other possibilities to produce blue profiles. The spectra located in the core mask are first averaged from the $2\times2$ pix$^2$ box and smoothed to a velocity resolution $\sim0.4$\,\kms. Then the spectra are overlaid on the green elliptical footprints of HCN-BP cores.

The mapping of three HCN cores BG023.968-00.110C1, BG029.397-00.095C1 and BG030.719-00.081C1 all perform various but coherent line profiles among the core. In other words, although the averaged spectrum over the core shows a significant blue profile, the individual spectra at different positions can change from blue to red profiles continuously. The variety can also be seen from the moment one (M1) maps in color map (Figure\,\ref{fig:specgrid}). The details of rotation axis calculation can be found Appendix\,\ref{app:rotate}. 

Finally, a total of 11 HCN-BP cores survive the ``mapping'' tests and provide a strong indication of infall motion. We also check whether there are central heating sources to build the temperature gradient in those massive clumps. Although it has the lowest luminosity-to-mass ratio of approximately 0.2 among the infrared dark clouds, BG028.565-00.236 still exhibits molecular outflows and H$_2$O/CH$_3$OH masers at higher angular resolution \citep{Lu2015G28}. These findings suggest the presence of active star formation and central heating sources, not to mention other sources with bright point-like or even extended infrared emission. The discussion in Section\,\ref{discuss:multi-J} will further strengthen the argument here, since the high-$J$ transition trace the denser (therefore inner) regions where the temperature gradient is guaranteed. Therefore, the blue profiles in the 11 HCN cores are most likely to be induced by infall motion. The subsample hereafter serve as promising candidates of infall in massive star-forming regions.


\section{Discussion} \label{sec:discuss}
\subsection{What Leads to Variety of Line Profiles at Multi-J Transitions?} \label{discuss:multi-J}

As demonstrated in Section\,\ref{result:profiles}, only 14 out of 38 clumps have blue profiles seen in \hcnft~lines, contributing to a profile retention rate of 36.8\% from low to high-level transitions (low-/high-$J$ for abbreviation where ``$J$'' represents the quantum number of the rotation transition). In addition, there are 4 other clumps with red profiles in \hcnft, while others have only one peaked or even no detection. Since all the clumps have evident blue profiles in \hcoptt, it is natural to ask what leads to the inconsistency of line profiles at dual-$J$ transitions.

We attribute the main factor for the inconsistency of profiles at multi-$J$ transitions to the difference of critical densities \footnote{Here, we use the same definition of critical density as \citet{Shirley2015Lines}: the critical density $n_{\rm crit}$ is defined as the molecular hydrogen density for which the net radiative decay rate from $j\to k$ equals the rate of collisional depopulation out of the upper level $j$ for a multi-level system.}. In our case, the critical density of \hcoptt~is $1.6\times10^6$\,cm$^{-3}$ at 10\,K and $1.4\times10^6$\,cm$^{-3}$ at 20\,K. On the other hand, the critical density of \hcnft~is $3.0\times10^7$\,cm$^{-3}$ at 10\,K and $2.3\times10^7$\,cm$^{-3}$ at 20\,K, which is approximately twenty times higher. Thus, different infall tracers, such as low-/high-$J$ transitions of HCO$^+$ or HCN species, should trace different parts or layers of dense star-forming clumps \citep{Xie2021HCOp}. As such, the infall profiles will be presented in the best way when the opacity of the source and the critical density of the tracer are well matched, as argued in \citet{Wu2003Infall}.

\subsubsection{Two Possible Scenarios} \label{multi-J:scenarios}

Given the different critical densities between \hcoptt~and \hcnft, there are two possible scenarios for our observed variety/inconsistency of line profiles at multi-$J$ transitions:

$\bullet$ While gas infalls in the outer envelope of massive clumps, the bulk motion can become more complex or even prohibited due to feedback from stars, such as outflows and stellar winds, or other dynamic processes occurring at a certain density layer. In some cases, the motion may even be reversed, resulting in an expanding motion. Consequently, there are multiple possibilities for bulk motion at the layer that \hcnft~traces, leading to a low detection rate of blue profiles at high-$J$ transitions. 

$\bullet$ The optical depth of molecular lines is determined primarily by the kinetic temperature $T_{\rm kin}$ and the column density of the molecule $N_{\rm mol}=N_{\rm H_2}\times X_{\rm mol}$, where $N_{\rm H_2}$ represents the column density of molecular hydrogen and $X_{\rm mol}$ denotes the abundance of the molecule. Due to variations in both $N_{\rm mol}$ and $T_{\rm kin}$ within our sample, the optical depth of the high-$J$ transition $\tau$(\hcnft) can vary significantly. Consequently, in some clumps, $\tau$(\hcnft) may not be sufficiently high to produce asymmetric line profiles, even if there is still gas infall motion present. 

\begin{figure}[!ht]
\centering
\includegraphics[width=1.0\linewidth]{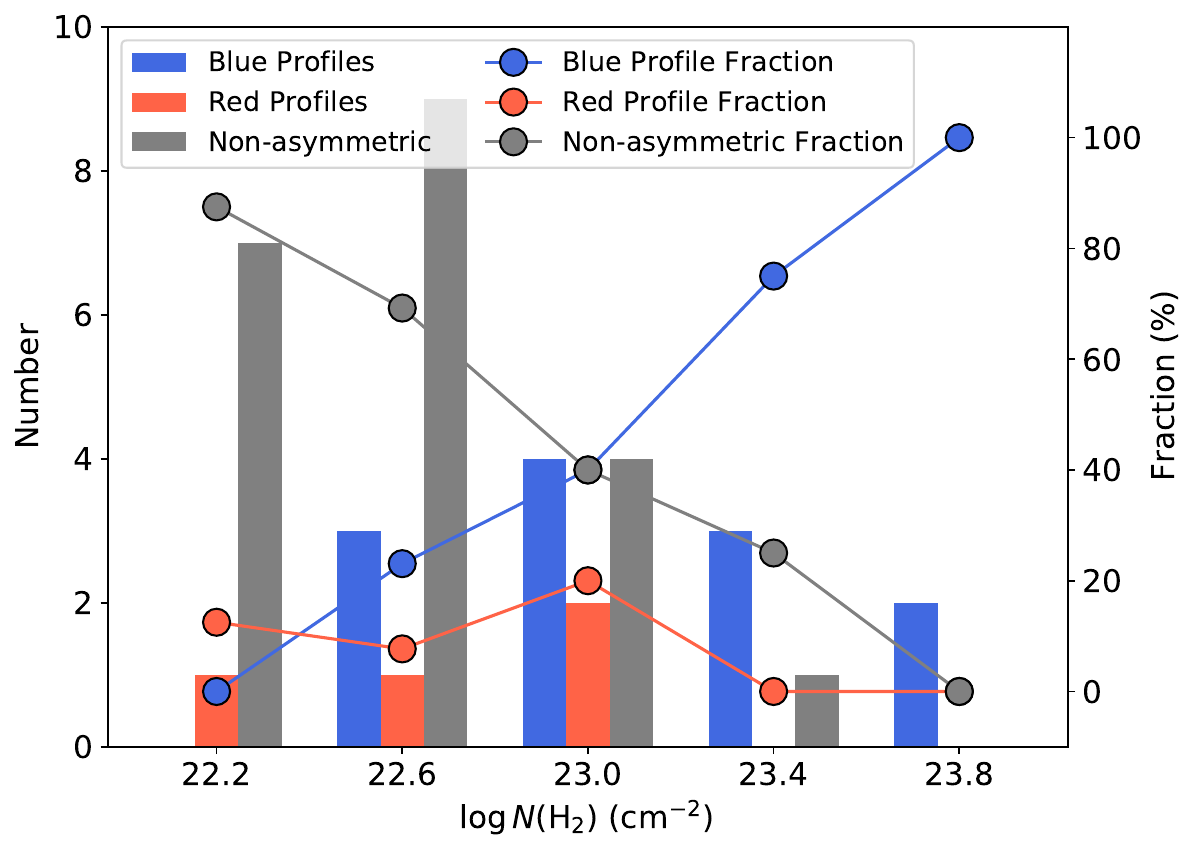}
\caption{The number distribution (histogram) and the fraction (line-connected scatter plot) of different line profiles change with peak column density bins $\log N(\rm H_2)\,(\rm cm^{-2})$ at the beam of 21\arcsec. The blue, red, and grey colors stand for blue profiles, red profiles, and non-asymmetric profiles (including single peaked and non-detection), respectively. \label{fig:fraction}}
\end{figure}

To distinguish between two scenarios, we can compare the predicted line profiles with the observed ones. For the first scenario, the fraction of distinct profiles should be determined by the likelihood of different types of bulk motions (infall, outflow/expansion, and static). For the second scenario, the detection rate of line profiles should be lower in clumps with lower column density, while the high-$J$ transition line should maintain the same profile as the low-$J$ transition line in clumps with a higher column density. Figure\,\ref{fig:fraction} displays that the fractions of both red profiles and non-asymmetric profiles systematically decrease while the fraction of blue profiles increases. Since all clumps have a blue profile \hcoptt, the rising trend of the fraction of \hcnft~with blue profiles suggests that the high-$J$ transition still conveys the same bulk motion information, but only in high-density clumps. 

\textit{Caveats}. We acknowledge that the peak column density $N_{\rm H_2}$ is based on an angular resolution of 21\arcsec, which is coarser than that of the JCMT. If the source has a centralized density distribution, the column density at the higher angular resolution should be higher than that at the lower angular resolution. Considering a Gaussian distribution of density and assuming that dust emission is optically thin, we can calculate how much column density is underestimated by $\mathcal{R}_{N}$,
\begin{equation}
\mathcal{R}_{N} = \frac{\iint_{\rm \Omega_{1}} \mathcal{G}(x,y;\sigma) \mathrm{d}\Omega}{\iint_{\rm \Omega_{2}} \mathcal{G}(x,y;\sigma) \mathrm{d}\Omega} - 1,
\end{equation}
where $\Omega_1$ and $\Omega_2$ are the JCMT and ATLASGAL beam solid angles respectively, $\mathcal{G}(x,y;\sigma)$ is a Gaussian density model with a dispersion of $\sigma$. For a typical value $\sigma=20\arcsec$ in our sample, $\mathcal{R}_N=0.31$, indicating that there can be a moderate systematic underestimation of column density if observed in the JCMT beam. But inversely speaking, we can smooth the JCMT lines into the same resolution of 21\arcsec, which is the same as that of column density. Since the profiles we discuss are from the averaged lines inside the cores where the profiles should be coherent (see Section\,\ref{result:profiles}), it’s safe to compare two in the context of 21\arcsec~resolution.

\subsubsection{Large Variations In Optical Depths of HCN (4-3)} \label{multi-J:tau}

\begin{figure*}[!ht]
\centering
\includegraphics[width=1.0\linewidth]{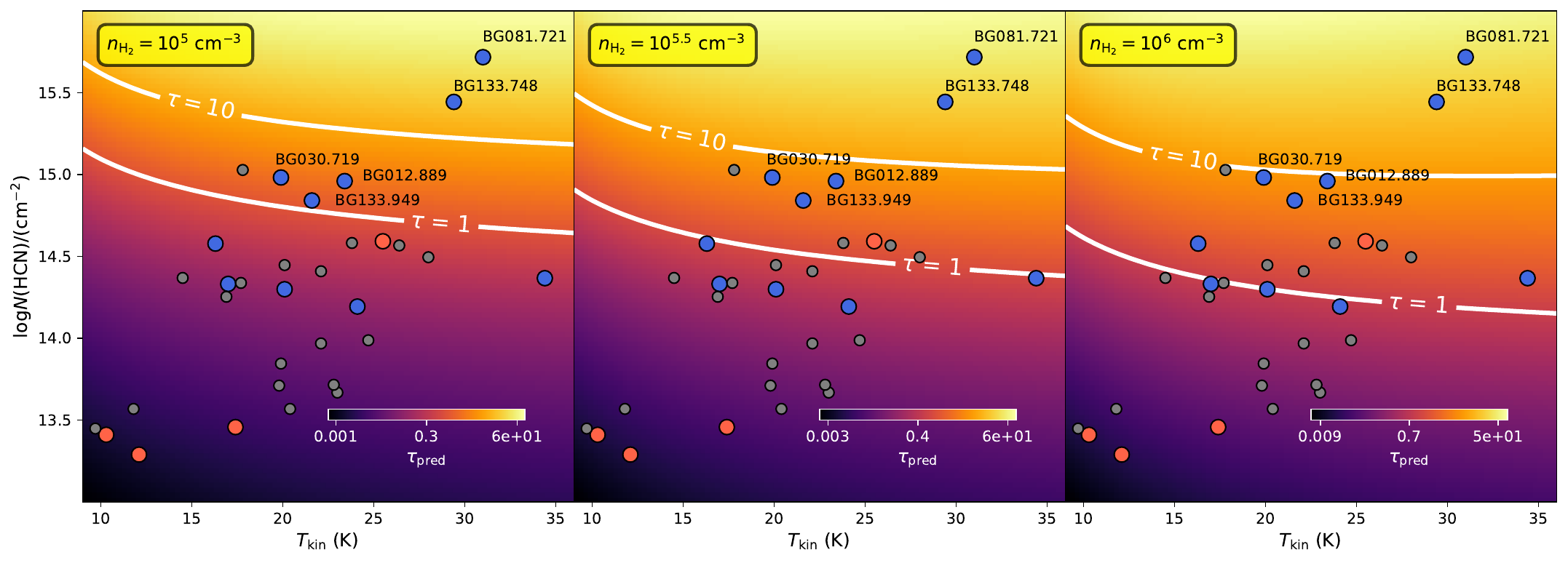}
\caption{The simulated grids of predicted optical depth ($\tau_{\rm pred}$) based on the kinetic temperature ($T_{\rm kin}$) and the column density of HCN ($N_{\rm HCN}$), with three different volume densities of collisional partner H$_2$ ($n_{\rm H_2}$). The white lines outline the contour levels of $\tau=$1 and 10. The 38 clumps are shown as filled circles, with blue, red, and gray colors representing blue profiles, red profiles, and non-asymmetric profiles, respecitvely. Five clumps with the highest $\tau_{\rm pred}$ are labeled. The colorbars are shown in the lower right. \label{fig:mockgrid}}
\end{figure*}

The optical depth of \hcnft~is calculated by \texttt{RADEX}\footnote{\href{https://personal.sron.nl/~vdtak/radex/index.shtml}{https://personal.sron.nl/~vdtak/radex/index.shtml}}, a computer program that calculates the strengths of the atomic and molecular lines of interstellar clouds, which are assumed to be homogeneous \citep{2007A&A...468..627V}. The presumed excitation conditions are: 1) background temperature of 2.73\,K; 2) collision partner (H$_2$) volume density of $10^{5}$--$10^{6}$\,cm$^{-3}$ in the HCN cores (see details in Appendix\,\ref{app:vden}; 3) \hcnft~line width of 8.8\,\kms, which is the mean value of observed spectra. We perform a $100\times100$ grid to calculate the optical depth of \hcnft, $\tau_{\rm pred}$, based on two variables, the kinetic temperature $T_{\rm kin}$ and the column density of the HCN molecule $N_{\rm HCN}$, within the parameter space defined by the observed values. 

The calculation grids are presented in Figure\,\ref{fig:mockgrid}, where contour levels of $\tau=$1 and 10 are represented by solid white lines. We utilize the sample of 38 observed massive clumps to predict the optical depths $\tau_{\rm pred}$. In our calculations, we assume that the kinetic temperature $T_{\rm kin}$ is equal to the dust temperature $T_{\rm dust}$. The column density of HCN $N_{\rm HCN}$ is determined using the relation $N_{\rm HCN} = N_{\rm H_2} \times X_{\rm HCN}$, where $N_{\rm H_2}$ is the H$_2$ column density and $X_{\rm HCN}$ is the abundance of HCN relative to H$_2$, which depends on the evolutionary stage \citep{Martinez2023abundance}. We assign different $X_{\rm HCN}$ values to clumps with distinct evolutionary types based on column (7) of Table\,\ref{tab:sample}. For type 0, $X_{\rm HCN}=5.6(1.1)\times10^{-10}$; for type 1, $X_{\rm HCN}=2.2(0.4)\times10^{-9}$; for type 2, $X_{\rm HCN}=5.9(0.3)\times10^{-9}$; for type 3, $X_{\rm HCN}=3.0(0.6)\times10^{-9}$. Additionally, considering the observed \hcnft~lines exhibit various velocity widths ($\mathrm{d}V_{\rm HCN}$) ranging from 5 to 15\,\kms, we calibrate the $N_{\rm HCN}$ values to account for the effect of $\mathrm{d}V_{\rm HCN}$ using Eq.\,3 in \citet{Remijan2004Lines}. 

Figure\,\ref{fig:mockgrid} illustrates that the optical depths of clumps exhibiting blue profiles ($\tau_{\rm B}$) consistently exceed those of non-asymmetric profiles ($\tau_{\rm NA}$) or red profiles ($\tau_{\rm RP}$). Furthermore, in the regime of $n_{\rm H_2}=10^6$\,cm$^{-3}$, all clumps demonstrate $\tau_{\rm B}\gtrsim1$, and even in the regime of $n_{\rm H_2}=10^5$\,cm$^{-3}$, the top five clumps (indicated with labels) maintain a high level of opacity. Conversely, for a significant proportion of clumps exhibiting red and non-asymmetric profiles, the optical depths do not exceed 1 under any given conditions. The outcome is in alignment with the findings presented in \citet{He2016Infall}, where the identified infall candidates exhibit elevated H$_2$ column densities and H$_2$ volume densities in contrast to the clumps where infall motions were not detected. We acknowledge that there are still several clumps, especially for BG013.882-00.143C1 with comparable $\tau_{\rm pred}$, but displaying a non-asymmetric profile or a red profile. In other words, even with enough optical depth, these clumps don't show blue profiles any longer. Therefore, the observations are likely to support hybrid scenarios, where an adequate optical depth is crucial for inducing blue profiles but the inner motion can also be complicated. Blue profile in low-$J$ transition should not guarantee blue profiles in high-$J$ transition. 

\textit{Caveats}. First, $T_{\rm kin}$ is assumed to be the dust temperature averaged through the clump $T_{\rm dust}$, which is a rough estimate. But as shown in Figure\,\ref{fig:mockgrid}, $T_{\rm kin}$ has much smaller variations in $\tau_{\rm pred}$ than $N_{\rm H_2}$ does. This suggests that the potential bias arising from the uncertainty in $T_{\rm kin}$ is mitigated. Second, there is no one-to-one correspondence between $X_{\rm HCN}$ and each individual source. Consequently, although these caveats result in a relatively rough estimation of $\tau_{\rm HCN (4-3)}$, the relative values of $\tau_{\rm HCN (4-3)}$ are reliable, allowing for qualitative analysis and further investigation.

\subsubsection{Triple-J Transition Lines In a Subsample} \label{multi-J:triple}

We cross match the 48 blue-profile clumps \citep[in \hcoptt~lines as reported by][]{Shirley2013FollowLine} with MALT90 surveys\footnote{The MALT90 survey is a large international project that exploited the fast-mapping capability of the ATNF Mopra 22-m telescope, \href{http://atoa.atnf.csiro.au/MALT90}{http://atoa.atnf.csiro.au/MALT90}} \citep{Jackson2013MALT90} of its low-$J$ transition counterpart \hcopoz~lines \citep[reported by][]{He2015Infall,He2016Infall}. Since the SMT and the Mopra are located in different hemispheres, we only have six sources overlapped with both the $J=1-0$ and $J=3-2$ transitions as a subsample. 

Table\,\ref{tab:multi-J} provides a compilation of line profiles from a subsample consisting of sources from \citet{Shirley2013FollowLine}, \citet{He2015Infall,He2016Infall}, and our work. Among the six sources with blue-profile \hcoptt~lines, four sources consistently exhibit blue-profile \hcopoz~lines, while the remaining two sources BG009.212-00.202 and BG012.889+01.480 display red-profile \hcopoz~lines. For the two sources with red-profile \hcopoz~lines, the classification in column (7) in Table\,\ref{tab:sample} indicates that they are both in a more evolved stage, which aligns with their extended infrared emission shown in Figure\,\ref{fig:infrared}. BG012.889+01.480 (also I18089-1732) was reported to contain a nearly face-on disk \citep{Sanhueza2021I18089} with collimated SiO\,(5-4) bipolar outflow \citep{Beuther2004I18089}. If the outflow direction is perpendicular to the disk plane, then the inclination angle of the outflow axis should be small and the outflow motion can provide enough expanding effects along the line of sight. The argument can be further verified in the case of BG009.212-00.202 by high-resolution observation. Once verified, it is likely that the red profiles are a result of outflows and bulk expanding expansion. Previous studies have demonstrated that \hcoptt~lines are capable of tracing infall motion in both the early \citep{Xie2021HCOp} and late stages of massive star-forming regions \citep{Fuller2005Infall,Reiter2011Infall,Klaassen2012Infall}. However, our work, although based on a limited sample size, suggests that low-$J$ transitions such as \hcopoz~may be more susceptible to contamination from other bulk motions present in the outer low-density layers. On the other hand, high-$J$ transitions like \hcoptt~appear to be more reliable for tracing infall motion in a more evolved stage of high-mass star-forming regions.

For the four sources with \hcnft~observations, we calculate the optical depths of \hcnft~lines based on the input parameters including column density of HCN ($N_{\rm HCN}$), kinetic temperature ($T_{\rm kin}$), line width ($\mathrm{d}V_{\rm HCN}$), and volume density of collision partner ($n_{\rm H_2}$). The first three parameters are directly retrieved from Table\,\ref{tab:sample} and \ref{tab:HCNpar}, while the last one $n_{\rm H_2}$ is given a range of ($10^5$, $10^{5.5}$ and $10^6$, cm$^{-3}$), which is similar to what has been done in Figure\,\ref{fig:mockgrid}. Overall, higher $n_{\rm H_2}$ results in higher level of thermalization due to collision process, and then result in higher excitation temperature $T_{\rm ex}$ and $\tau_{\rm HCN (4-3)}$ until the critical density of $n_{\rm crit, HCN(4-3)}$ of $2.3\times10^7$\,cm$^{-3}$. Although $\tau_{\rm HCN (4-3)}$ varies by magnitude along with $n_{\rm H_2}$, BG008.458-00.222 and BG011.083-00.536 with single-peaked profiles have $\tau_{\rm HCN (4-3)}\ll1$; BG009.212-00.202 and BG012.889+01.480 with blue profiles always have $\tau_{\rm HCN (4-3)}\gtrsim1$. Therefore, the optical depth is the main reason for the variations of line profiles in \hcnft.

However, we still note that the conclusion is not significant because of the limited sample size. Therefore, it's encouraging to survey multi-$J$ transitions in a much larger and less biased sample and check whether the conclusion holds or not.

\subsubsection{Low Detection Rate of Blue Profiles and Their Connection to Infall Motions} \label{multi-J:lowDR}

\begin{figure}[!ht]
\includegraphics[width=1.0\linewidth]{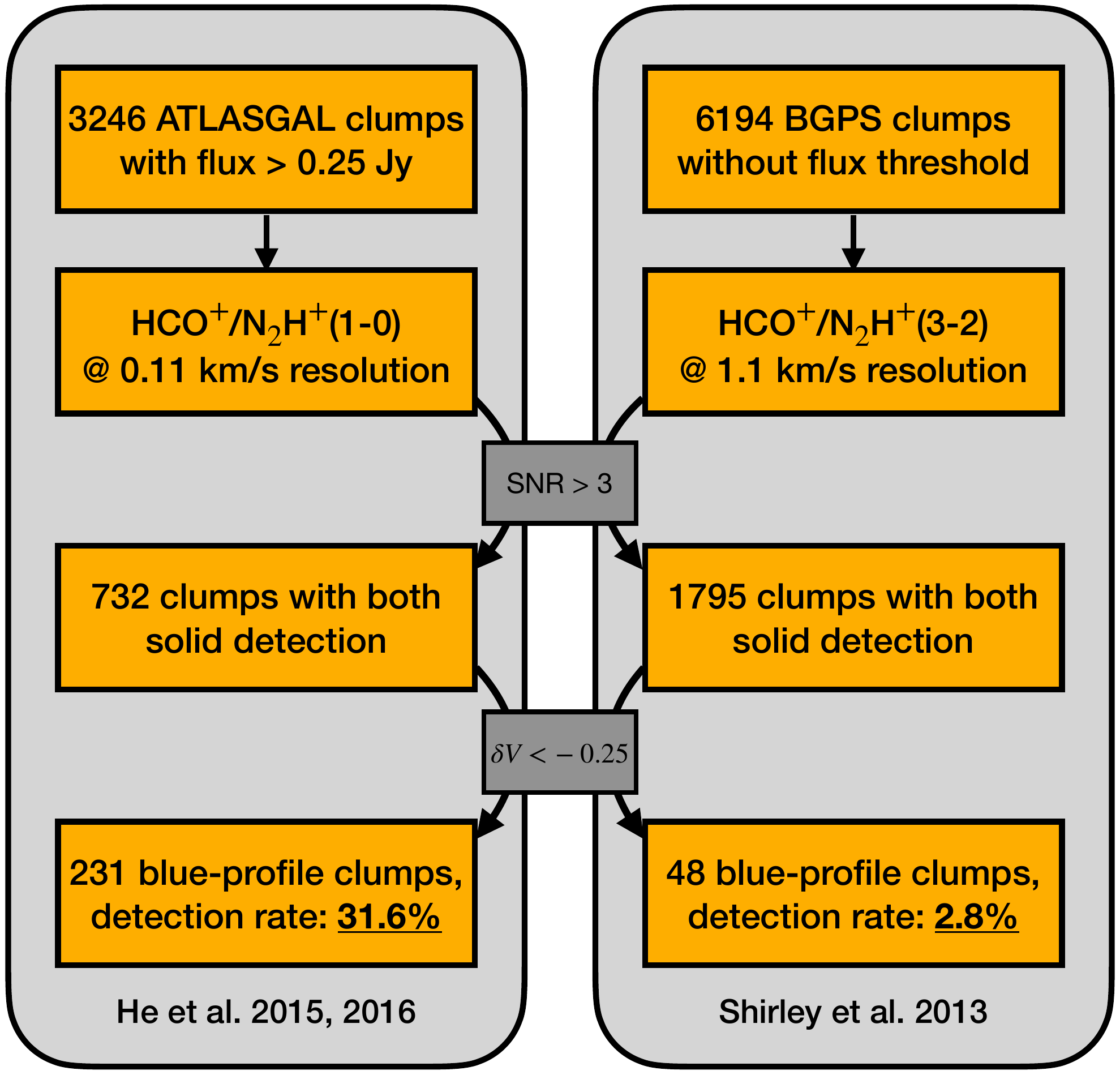}
\caption{The low detection rate of blue profiles in star forming clumps guided by ATLASGAL and BGPS. 3246 ATLASGAL clumps with 870\,$\mu$m flux larger 0.25\,Jy are observed in HCO$^+$/N$_2$H$^+$\,(1-0) at a spectral resolution of 0.11\,\kms, by the Mopra MALT90 line survey. 732 clumps have solid detection of SNR$>$3, among which 231 clumps have blue profiles by canonical criterion of $\delta V<-0.25$. The detection rate of blue profile is then 31.6\% \citep{He2015Infall,He2016Infall}. 6194 BGPS clumps without flux threshold are observed in HCO$^+$/N$_2$H$^+$\,(3-2) at a spectral resolution of 1.1\,\kms, by the SMT line survey \citep{Schlingman2011FollowLine,Shirley2013FollowLine}. 1795 clumps have solid detection of SNR$>$3, among which 48 have blue profiles, donating a detection rate of 2.8\% \citep{Shirley2013FollowLine}. \label{fig:lowDR}}
\end{figure}

\begin{figure*}[!ht]
\centering
\includegraphics[width=0.48\linewidth]{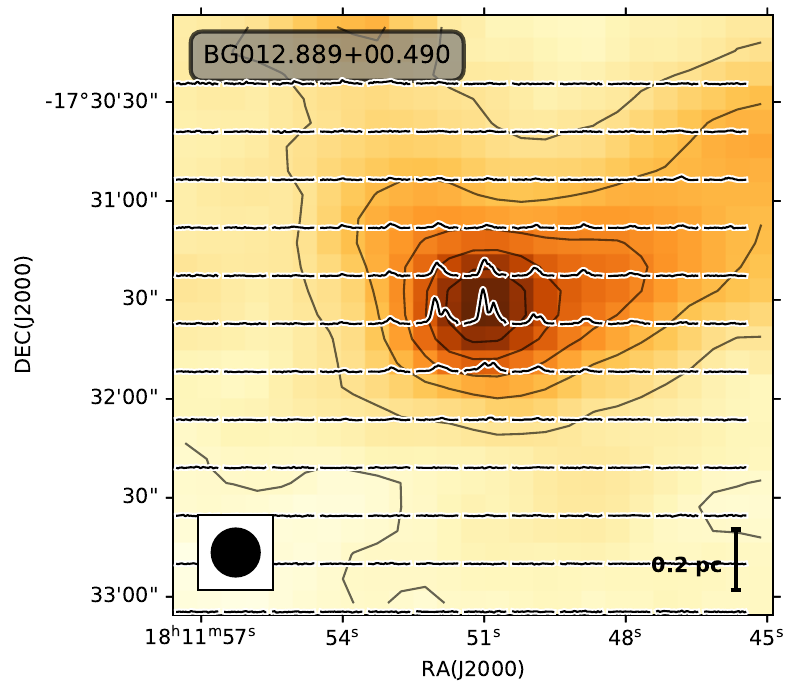}
\includegraphics[width=0.48\linewidth]{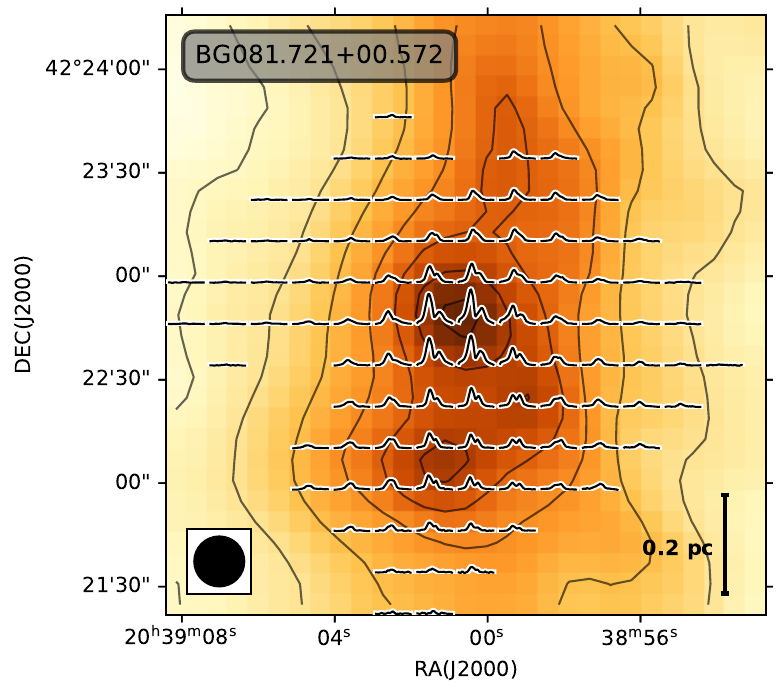}
\includegraphics[width=0.48\linewidth]{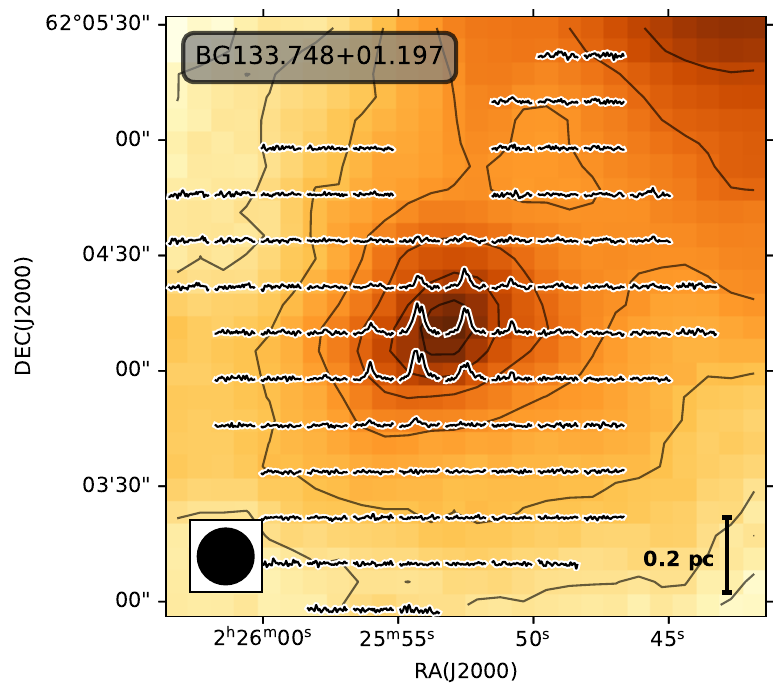}
\includegraphics[width=0.48\linewidth]{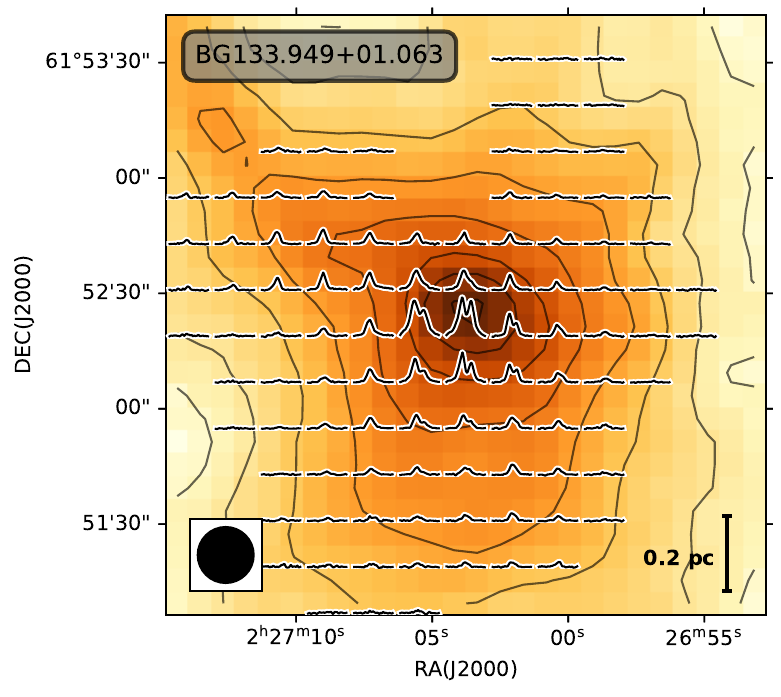}
\caption{The grid (2$\times$2 pix$^2$) of \hcnft~line profiles overlaid on the continuum map (either ATLASGAL or SCUBA-2) for BG012.889, BG081.721, BG133.748, and BG133.949, respectively. The beam size of HARP receiver at 350\,GHz is shown on the bottom left. The scale bar of 0.2\,pc is shown on the bottom right. \label{fig:profilemap}}
\end{figure*}

As summarized in Figure \ref{fig:lowDR}, two systematic investigations have been undertaken to identify blue profiles within massive star-forming clumps, employing the ATLASGAL and BGPS follow-up line surveys, respectively. However, the detection rate of the blue profiles in BGPS clumps is found to be more than ten times lower than that observed in ATLASGAL clumps. This substantial discrepancy can be attributed to two primary factors. 

Firstly, ATLASGAL clumps are observed using low-$J$ transitions, whereas the BGPS clumps employ high-$J$ transitions. Furthermore, it is important to consider that the ATLASGAL line survey implemented a flux threshold of 0.25\,Jy at 870 $\mu$m. This threshold was applied to ensure the inclusion of clumps with a mass of 200 M$_\odot$, assuming a distance of 10\,kpc and a temperature of 10\,K \citep{Jackson2013MALT90}. In contrast, the BGPS line survey did not impose any specific flux threshold. Consequently, the absence of a threshold in the BGPS survey results in the inclusion of a broader range of clumps, including those with lower flux values. Consequently, the opacity of the high-$J$ transition line, especially in low-flux clumps, may not be sufficiently high to produce the characteristic self-absorption signature, as discussed in Section \ref{multi-J:tau}, therefore diluting the overall detection rate of blue profiles in the BGPS sample. 
Second, the MALT90 line survey has a spectral resolution that is ten times greater than that of the BGPS line survey. It should be noted that a low spectral resolution is inadequate for detecting blue profiles induced by low infall velocities. These factors contribute significantly to the disparity observed in the detection rates of blue profiles between the two surveys. Moreover, such a comparison of detection rate also encourages further investigations with enhanced spectral resolution and the use of appropriate transition lines. 

\begin{figure*}[!ht]
\centering
\includegraphics[width=0.32\linewidth]{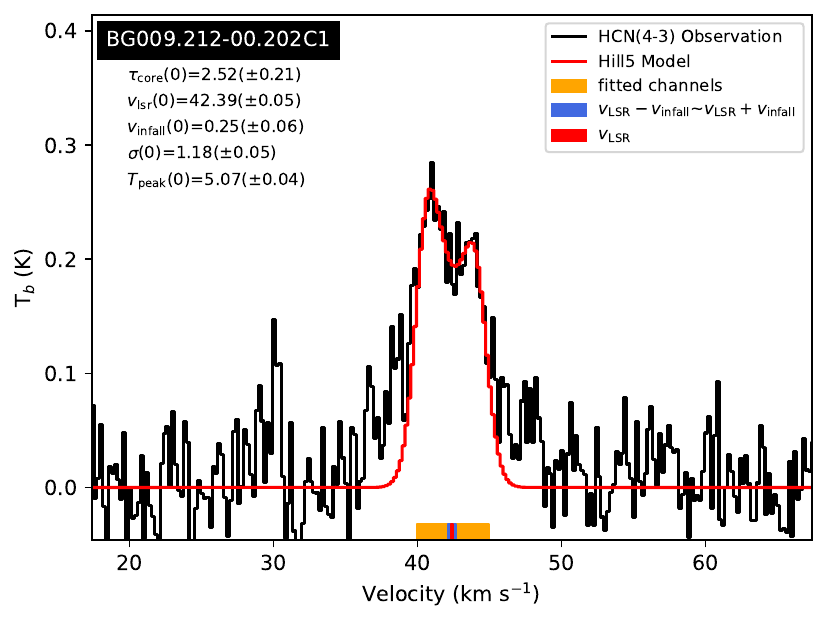}
\includegraphics[width=0.32\linewidth]{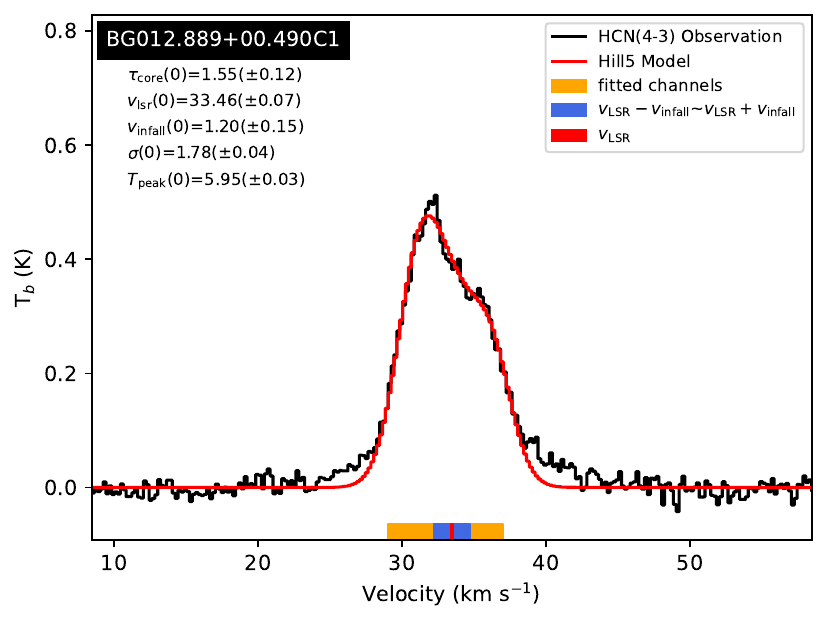}
\includegraphics[width=0.32\linewidth]{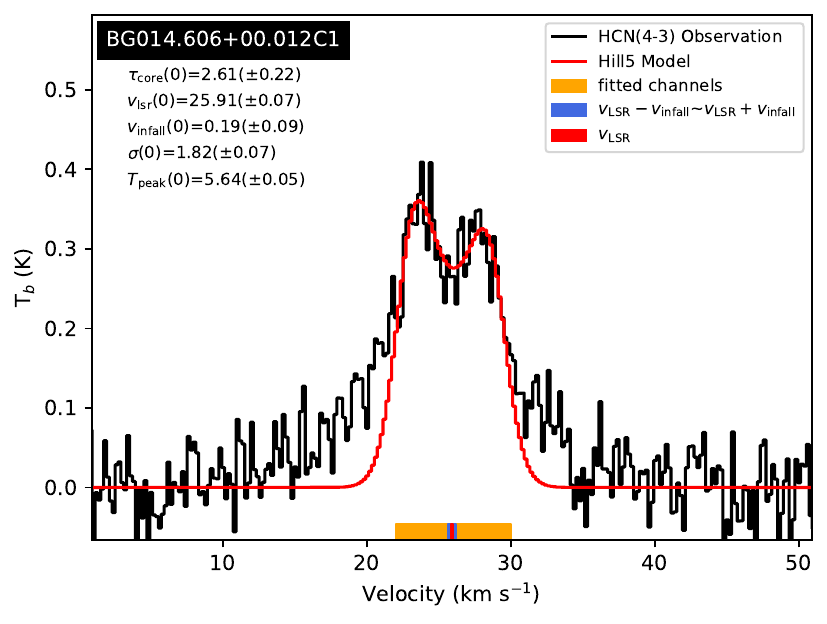}
\includegraphics[width=0.32\linewidth]{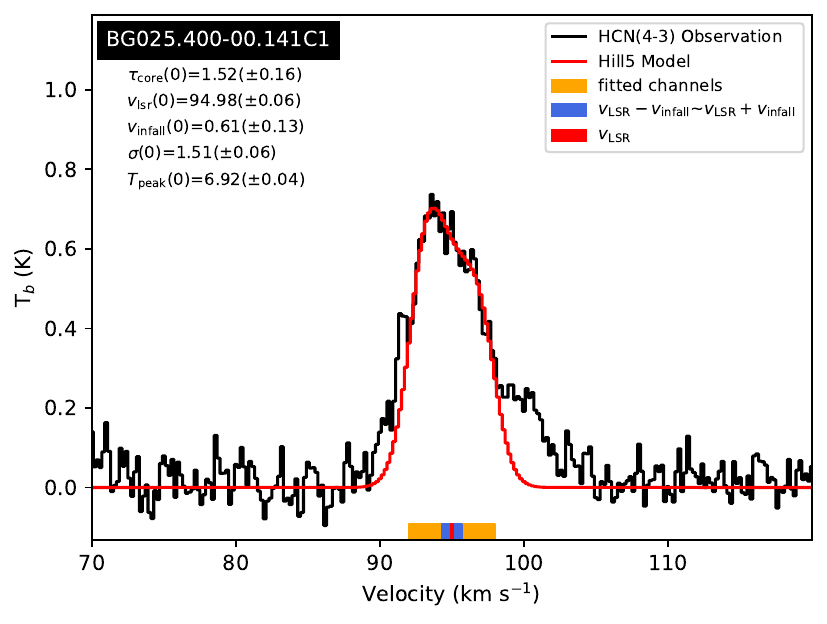}
\includegraphics[width=0.32\linewidth]{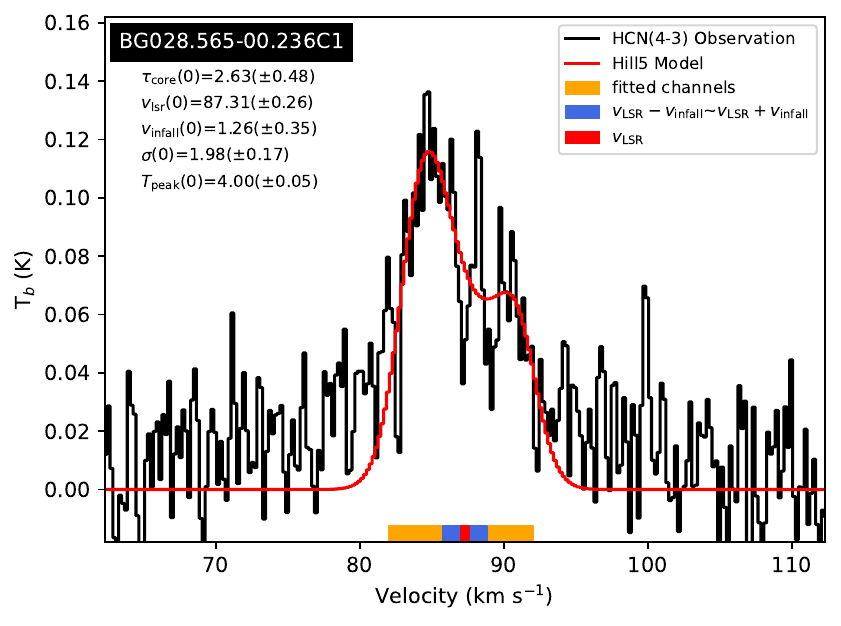}
\includegraphics[width=0.32\linewidth]{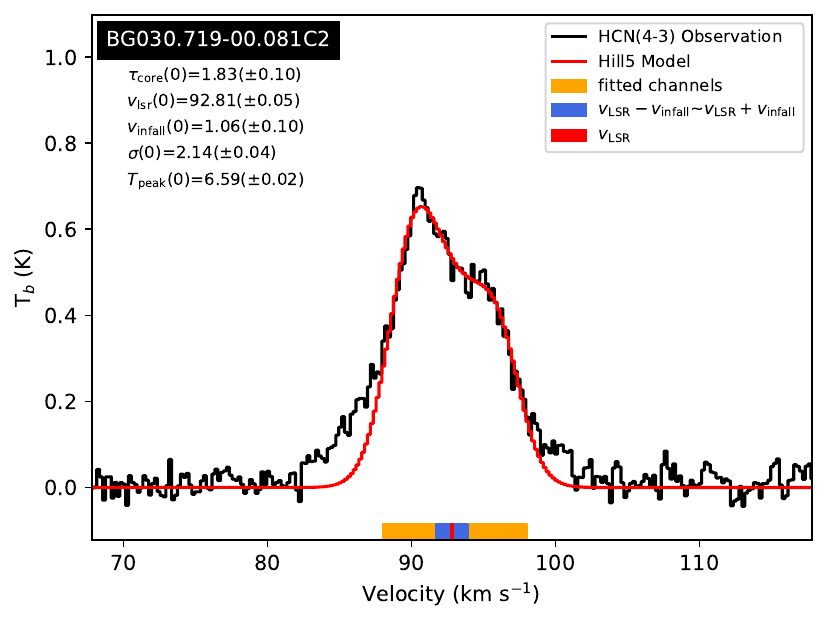}
\includegraphics[width=0.32\linewidth]{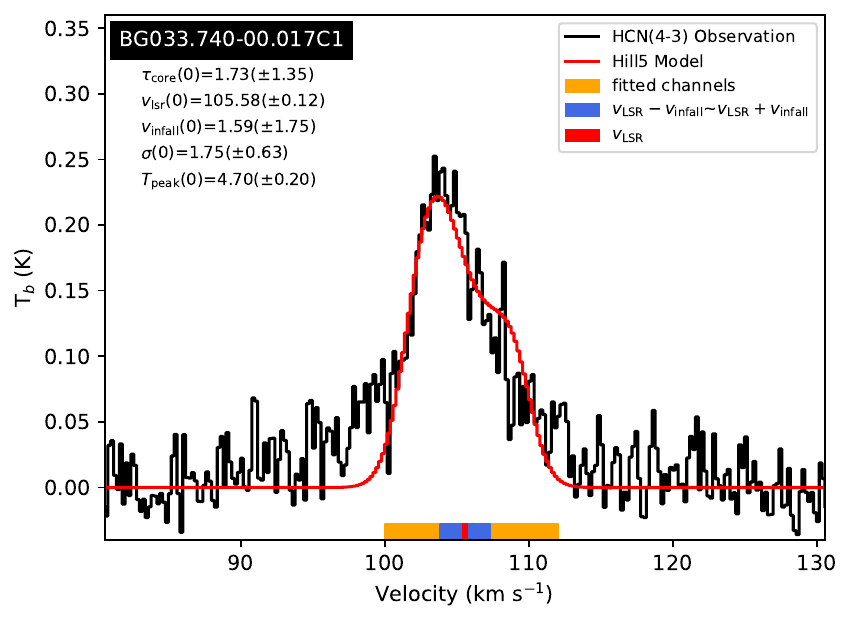}
\includegraphics[width=0.32\linewidth]{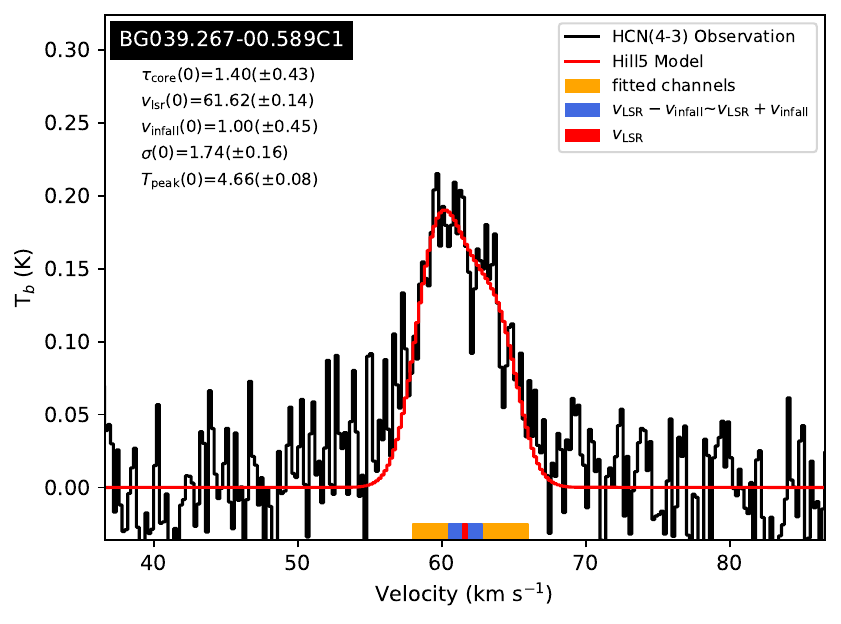}
\includegraphics[width=0.32\linewidth]{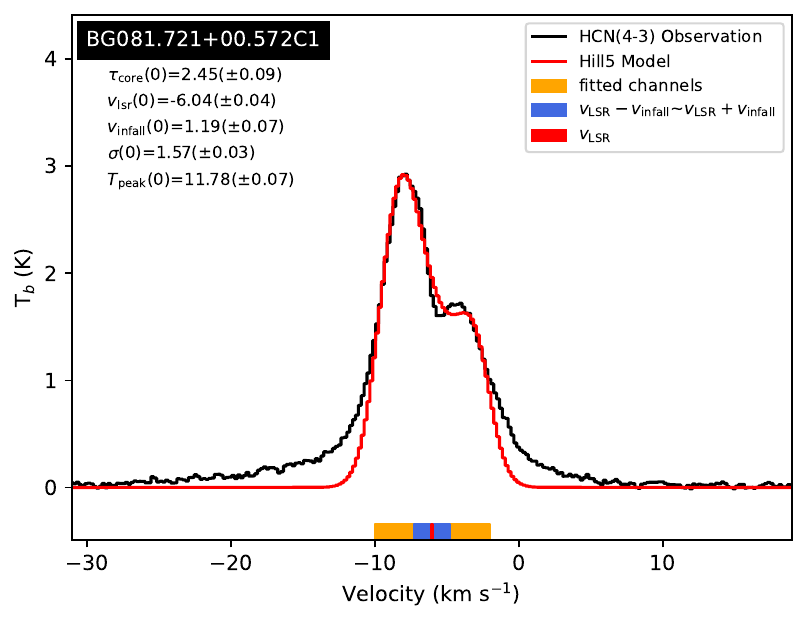}
\includegraphics[width=0.32\linewidth]{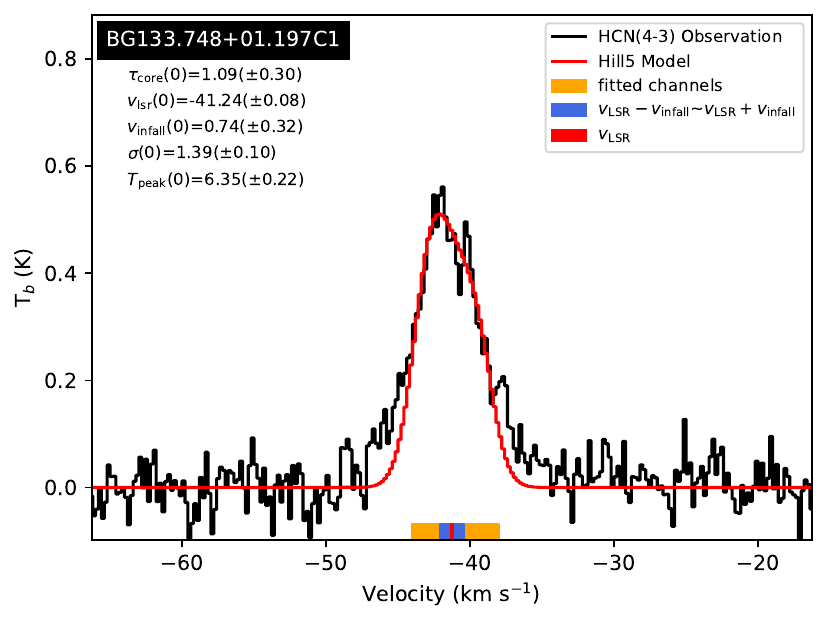}
\includegraphics[width=0.32\linewidth]{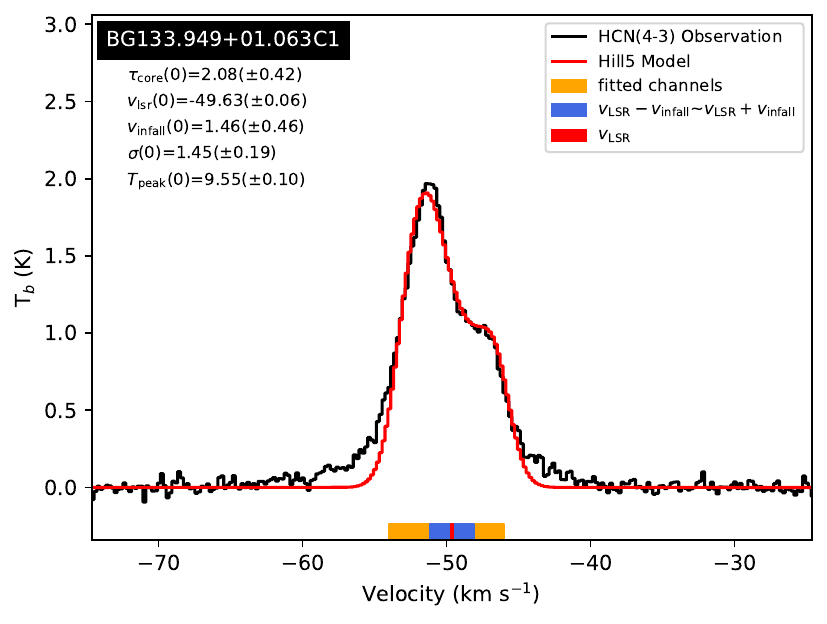}
\caption{The averaged \hcnft~lines are indicated in black solid line while the best-fit Hill5 model are in red solid line, with five parameters shown on the upper left corner. We highlight the velocity range used for Hill5 model fitting in orange color at the bottom. The red band indicates the systematic velocity $v_{\rm LSR}$. The blue band indicates the velocity span of the infall motion, that is, from $v_{\rm LSR}-v_{\rm infall}$ to $v_{\rm LSR}+v_{\rm infall}$. \label{fig:Hill5_Fitter}}
\end{figure*}

Nevertheless, it's essential to recognize that studying blue profile is a phenomenology after all, and establishing a direct link between such profiles and infall motions remains a challenging endeavor. Despite advancements in observational capabilities, the comprehension of blue profiles is impeded by limited insights into the physical conditions prevailing within these regions, such as the distribution of temperature and density. Additionally, the intricate nature of infall motions within these clumps, coupled with potential influences from feedback mechanisms, contributes to a notable false positive rate when inferring infall motion from blue profiles. To disentangle the intricate implications of blue profiles, investigations of high-resolution are imperative, encompassing thorough analyses of gas kinematics \citep[refer to Section 4.4 in][]{Xu2023SDC335}.

On another front, systematic examinations of blue profiles also grapple with a significant false negative rate. For example, the low detection rate of blue profiles in BGPS clumps does not necessarily imply a low detection rate of infall motion, underscoring the importance of using suitable tracers for identifying infall motion. A careful balance must be maintained between the detectability of blue profiles and the ability of a tracer to probe the desired depth. Low-$J$ transitions, for instance, tend to produce blue profiles more readily due to sufficient opacity, albeit primarily tracing the gas envelope. On the other side, high-$J$ transitions effectively capture inner gas motion but may be too optically thin to generate blue profiles in clumps with low column densities.

\subsection{Mapping Infall Motions in Massive Star-forming Clumps} \label{discuss:infall}

\subsubsection{Mapping Clump-scale Global Collapse} \label{discuss:globalcollapse}

JCMT mapping observations with an angular resolution of 14\arcsec~have provided valuable insights into the resolution of infall motions. Among the observed clumps with \hcnft~blue profiles, there are five clumps with the highest predicted optical depths ($\tau_{\rm pred}$), namely BG081.721, BG133.748, BG030.719, BG012.889, and BG133.949.
It is worth noting that these five clumps also exhibit consistently highest SNRs as illustrated in Figure\,\ref{fig:Gaussian_Fitter}, suggesting the robustness of \texttt{RADEX} line simulation. Furthermore, they show the most extended emission patterns, as depicted in Figure\,\ref{fig:specgrid}, further emphasizing their significance in the study of infall motion. However, it should be noted that BG030.719 presents a unique case in which two separated HCN cores with opposite profiles are observed, resulting in a limited number of pixels available for mapping the infall motion with sufficient sampling. Therefore, excluding BG030.719, the remaining four clumps serve as a subsample that can be effectively utilized for infall motion mapping analyses.

As shown in all four clumps in Figure\,\ref{fig:profilemap}, the \hcnft~lines show strong spatial correlation with submillimeter continuum (dust) emission and blue-profile spectra in most of the clumps. Such line profiles are expected for an optically thick tracer of idealized collapsing clouds in which the excitation temperature is rising towards the center. What is important to note here is the extent (over at least 9 independent beams) over which this spectral signature is observed, and the absence of any other line asymmetries (cf. Section\,\ref{result:candidate}), strongly suggesting that all the four clumps are undergoing global collapse \citep[see a typical example of SDC335 in][]{Peretto2013SDC335,Xu2023SDC335}. A radiation transfer model combined with temperature and density profile derived from the far-infrared data can be used to fit the map of line profiles, inferring the infall velocity and mass infall rate (Xie et al. in preparation).

\subsubsection{Infall Parameters Fitted by Hill5 Model} \label{infall:hill5}

According to Section\,\ref{result:candidate}, a sample of 11 HCN cores are undergoing infall motion. To estimate the infall velocity, we used the ``Hill5'' model first introduced by \citet{DV2005Hill5}. The ``Hill5'' model assumes that the excitation temperature in the front of the cloud increases inward as a linear function of optical depth. Compared to the traditional ``two-layer'' model where excitation temperature is constant, ``Hill5'' model is more suitable for real physical scenarios where young stellar objects (YSOs) heat the core from inside out. Besides, \citet{DV2005Hill5} demonstrate that two-peak profiles are best matched by the ``Hill5'' model while ``Hill5'' and ``two-layer'' perform equally well for red-shoulder blue profiles without double peaks (red-shoulder hereafter). The ten infall candidates identified in Section\,\ref{result:candidate} are either double-peak or red-shoulder blue profiles, so the ``Hill5'' model is a better choice.

The model has five free parameters to fit: (1) the peak excitation temperature $T_{\rm peak}$, (2) the velocity dispersion of the molecular line $\sigma$, (3) the optical depth of the core $\tau_{\rm core}$, (4) the systematic velocity $v_{\rm LSR}$, and (5) the infall velocity of the gas in the core $v_{\rm infall}$. Formula derivations of the model are presented in detail in Appendix\,\ref{app:hill5}.

To determine the global accretion rate towards these cores, we fit the average spectra of the \hcnft~emission across the cores. The signal-to-noise ratio of the ten spectra has an average of $\sim30$, satisfying the criterion for the ``Hill5'' model fitting. Although two spectra of BG028.565-00.236C1 and BG039.267-00.589C1 have relatively lower $\mathrm{SNR}\gtrsim6$, the high spectral resolution of 0.2\,\kms~assure enough effective data points for model fitting. Most cores have extended velocity wings that are assumed to be induced by molecular outflows. To reduce the contamination from the wings, we cut out the velocity channels which are non-Gaussian parts in Gaussian fitting process (see Section\,\ref{result:avgspectra}). The preserved channels are fitted channels marked as orange bands in Figure\,\ref{fig:Hill5_Fitter}, with a bandwidth of 30--60 channels to cover the blue profile features. The uncertainties in the fitting are given by Python package \texttt{lmfit} that explicitly explore the parameter space and determine confidence levels. As initial guesses of the fit we assume $\tau_{\rm core}$ ranging from 0.1 to 30, a $v_{\rm LSR}$ between $v_{\rm LSR}-5$\,\kms~and $v_{\rm LSR}+5$\,\kms, $v_{\rm infall}$ between 0.1 and 4\,\kms, $\sigma$ between 0 and $\sigma_{\rm HCN}$ (from Gaussian fitting), and $T_{\rm peak}$ between 2.73 and 30\,K.

The fitted spectra are shown in Figure\,\ref{fig:Hill5_Fitter} with five parameters shown on the top left of each panel. The blue band indicates the velocity range of the infall motion, that is, from $v_{\rm LSR}-v_{\rm infall}$ to $v_{\rm LSR}+v_{\rm infall}$. The fitted parameters as well as the velocity channels used for fitting are sorted in columns (3)--(7) of Table\,\ref{tab:Hill5}.

\subsubsection{Infall Velocity vs. Free-fall Velocity} \label{infall:freefall} 

Since the free-fall velocity represents the typical timescale of gravitational collapse of a star-forming clump, comparing observed infall velocity to free-fall velocity helps to understand how fast the star formation proceeds in these massive clumps. The free-fall velocity $v_{\rm ff}$ is calculated by:
\begin{equation} \label{eq:ff}
    v_{\rm ff} = \sqrt{\frac{2GM_{\rm enc}}{R}},
\end{equation}
where $M_{\rm enc}$ is the mass enclosed within radius of $R$. Since the HCN cores are mostly even smaller than the clumps, we need to scale down $v_{\rm ff}$ at the HCN core scale. Substituting Eq.\,\ref{eq:M-R} into Eq.\,\ref{eq:ff}, free-fall velocity should be constant over the self-gravitating clumps. Therefore, we can directly compare the infall velocity and the free-fall velocity of the HCN cores. As shown in column (5) of Table\,\ref{tab:Hill5}, the infall velocity of the 11 clumps has a range of 0.2--1.6\,\kms, with mean and median values of 1.0 and 1.1\,\kms. Adopting the clump radius and mass in column (8) and (11) of Table\,\ref{tab:sample} into Eq.\,\ref{eq:ff}, the free-fall velocity has a range of 2.0--6.8\,\kms, with mean and median values of 3.6 and 3.2\,\kms. Therefore, the infall velocity fraction $\mathcal{F}_{\rm infall}$, defined as the ratio of the infall velocity to free-fall velocity, ranges from 5\% to 74\%, with both mean and median values of 32\%. The minimum value is consistent with what has been found in \citet{Wyrowski2016Infall}, but the maximum and mean/median values are systematically larger. However, the large fraction should be due to the different distances in our sample, because the radii of clumps with smaller distance tend to have lower mass ($M_{\rm clump} \propto D^2$ where $D$ is the distance). If we exclude the three nearest clumps BG081.721+00.57, BG133.748+01.19, and BG133.949+01.06, then the median fraction is $\sim20$\%.

Since the timescale is directly related to the velocity at a given radius, the ratio of the infall timescale ($\tau_{\rm infall}\propto 1/v_{\rm infall}$) to the free-fall timescale ($\tau_{\rm ff}\propto 1/v_{\rm ff}$) is inversely proportional to $\mathcal{F}_{\rm infall}$. This means that the dense region of the clump, as indicated by the HCN cores, will undergo collapse within a few to several tens of free-fall timescales.

\subsubsection{Mass Infall Rate} \label{infall:rate}

Assuming a simplified spherical model, the mass infall rate is calculated by \citep{Lopez2010Infall},
\begin{equation}
\begin{split}
\dot M & = 4\pi R^2 \rho v_{\rm infall} = 4\pi m_{\rm p} \mu_{\rm H_2} n_{\rm H_2} R^2 v_{\rm infall} \\
& = 8.9\times10^{-4} \left(\frac{\mu_{\rm H_2}}{2.809}\right) \left(\frac{n_{\rm H_2}}{10^{5}\,\mathrm{cm}^{-3}}\right) \left(\frac{R_{\rm deconv}}{0.1\,\mathrm{pc}}\right)^2 \\ 
& \left(\frac{v_{\rm infall}}{1\,\mathrm{km}\,\mathrm{s}^{-1}}\right) M_{\odot}\,\mathrm{yr}^{-1},
\end{split} 
\end{equation}
where $\mu_{\rm H_2}$ is the molecular weight per hydrogen molecule \citep[$\mu_{\rm H_2}=2.809;$][]{Evans2022SlowSF}, $n_{\rm H_2}$ and $R_{\rm deconv}$ are the volume density and physical radius of the defined HCN core, and $v_{\rm infall}$ is the infall velocity fitted from ``Hill5'' model. The $n_{\rm H_2}$ is estimated by $N_{\rm H_2}/(2 R_{\rm deconv})$. We note that the HCN core BG039.267-00.589C1 has not been resolved so we use the physical scale of the beam size as an upper limit, and therefore the derived mass infall rate is an upper limit as well. The calculated mass infall rate is then listed in Column (8) of Table\,\ref{tab:Hill5}. 

The mass infall rate exhibits a wide range, ranging from 0.15 to 32.1 $\times 10^{-3}$ \massrate, which aligns with typical values observed in high-mass clumps \citet{Yu2022InfallCat,He2016Infall}. The mean and median values of the mass infall rate are $7.6 \times 10^{-3}$ and $4.5 \times 10^{-3}$\,\massrate, respectively, which are in good agreement with the values derived from \hcopoz~lines in a sample of 11 IRDCs \citep{Xie2021HCOp} and HCO$^+$/HNC\,(1-0) lines in a sample of 33 IRDCs \citep{Pillari2023HCOp}. It should be noted that the mass infall rate obtained from \hcnft~lines primarily traces the inner regions of massive clumps, while the mass infall rate derived from \hcopoz~lines predominantly represents the outer parts or envelopes. Nevertheless, the remarkable consistency in mass infall rates between these two tracers suggests a continuous accretion process from the clump envelope to the inner region. This finding is supported by previous studies indicating minimal variations in mass infall rates during the evolution of high-mass clumps \citep{He2016Infall}. Utilizing multi-$J$ comparisons allows us to establish a connection between mass infall rates at various scales. Therefore, it is crucial to conduct follow-up high-resolution observations (e.g., by ALMA, NOEMA, or SMA) to precisely quantify the amount of mass ultimately transferred to the protostars. For instance, \citet{Xu2023SDC335} collected three-scale observations and revealed a consistent accretion process from clump-scale global collapse to core-scale gas feeding in the case of SDC335. Furthermore, high-resolution observations can improve our understanding of the concept of global collapse. Although a spherical model featuring a collapsing shell can adequately explain the blue profiles, most observations indicate that the inflows manifest as gas streams or elongated filamentary structures \citep{Peretto2013SDC335,Kirk2013FAccretion,Lu2018Filament,Liu2016EWBO,Xu2023SDC335,Yang2023HFS}. The ongoing SMA observations of six of our samples are promising for deepening our understanding of the inner dense gas distribution and kinematics as well.

\section{Conclusions} \label{sec:conclude}

Leveraging the efficient-mapping advantages of the JCMT HARP instrument, we perform an \hcnft~mapping survey of 38 representative massive star-forming clumps in the Bolocam Galactic Plane Survey (BGPS) guided with \hcoptt~``blue asymmetric line profile'' (blue profile). The high-$J$ transition with critical density of $>10^7$\,cm$^{-3}$, combined with previous low-$J$ transition data, mapping observational mode, and a wide range of physical properties in such a large sample, help deepen our understanding of blue profiles and their connection to gas infall motion in massive star-forming clumps. Our main findings are summarized as follow.

\begin{enumerate}
    \item We integrate the line intensity of \hcnft~lines and produce 38 \hcnft~moment 0 (M0) maps, of which 32 have detection and six have no detection. 30 M0 maps show isolated emission regions while two M0 maps show double emission regions. In total, 34 HCN emission cores (HCN cores) are identified by the \SE~algorithm. \hcnft~spectra extracted from the HCN cores have consistent velocities with that of \nthptt~lines, justifying the usage of \nthptt~as the systematic velocity tracer. 

    \item The averaged \hcnft~lines rather show various line profiles, including 14 blue, 4 red, and 22 non-asymmetric profiles, rather than keeping the same blue profile as the lower-$J$ transition \hcoptt~performs. Adopting the \hcnft~maps, we found the intrinsic variations of the line profile in three HCN cores, suggesting potential rotation motion. The rest of 11 HCN cores serve as a promising candidates of infall motion in massive star-forming regions.

    \item We find an increasing rate of blue profiles along the H$_2$ column density and the opacity of \hcnft~lines calculated from the non-LTE radiation transfer code \texttt{RADEX}, suggesting insufficient opacity should be the main reason for the low profile retention rate of 36.8\% (14 blue profiles out of 38 massive clumps). However, even with sufficient \hcnft~opacity, there are still some detections of red or non-asymmetric profiles, which suggest gas undergoing different motion at different density layers, traced by different transitions.

    \item A six-source subsample has three transitions, \hcopoz, \hcoptt, and \hcnft, with critical density ranging from $4.5\times10^4$\,cm$^{-3}$ to $2.3\times10^7$\,cm$^{-3}$. Although limited by sample size, single peaked line profiles systematically have low opacity $\tau\ll 1$ while blue profiles have enough high opacity $\tau\gtrsim 1$. Additionally, we find that two sources, namely BG009.212-00.202 and BG012.889+00.490, which exhibit bipolar outflows at relatively small inclination angles, display red profiles in the lowest-$J$ transition of \hcopoz. These profiles can be attributed to the presence of expanding gas envelopes along the line of sight.

    \item Comparison between two line surveys guided by ATLASGAL \citep{He2015Infall,He2016Infall} and BGPS \citep{Schlingman2011FollowLine,Shirley2013FollowLine} highlights the importance of appropriate tracer, high spectral resolution, and column density threshold of searching for blue profiles in a large sample. We also caution that the blue profile is a phenomenology after all, and the connection between the blue profile and the infall remains to be calibrated by a multi-$J$ transition line survey for a large sample.

    \item If all 11 blue profiles are produced by infall motions, we adopt the ``Hill5'' model to fit the infall velocity of the HCN cores, ranging from 0.2 to 1.6\,\kms, with mean and median value of 1.0 and 1.1\,\kms. The infall velocities account for a fraction of 5\% to 74\% to free-fall velocity, indicating the HCN cores will collapse within a few to several tens of free-fall timescales. 
    
    \item Assuming a simplified spherical model, the mass infall rate can be calculated, ranging from 0.15 to 32.1$\times10^{-3}$\,\massrate, with mean and median values of $7.6 \times 10^{-3}$ and $4.5 \times 10^{-3}$\,\massrate, which is consistent with what has been found in the low-$J$ transition \hcopoz. The consistency of the mass infall rate among different transitions (i.e., difference density layers) suggests a steady accretion process from the clump gas envelope to the inner region, as proposed by \citet{Xu2023SDC335}.
\end{enumerate}

\clearpage


\begin{longrotatetable}
\begin{deluxetable*}{cccccccccccccc}
\tabletypesize{\small}
\tablewidth{0pt}
\linespread{1.1}
\tablecaption{Sample and JCMT HARP HCN (4-3) Observations \label{tab:sample}}
\tablehead{
\colhead{JCMT Field} & \multicolumn{2}{c}{Position} & \colhead{$V_\mathrm{LSR}$} & \multicolumn{2}{c}{Distance} & \colhead{Type\tablenotemark{b}} & \colhead{$R_{\rm cl}$} & \colhead{$T_{\rm d}$} & \colhead{$\log(L_{\rm bol})$} & \colhead{$\log(M_{\rm cl})$} & \colhead{$\log(N_{\rm H_2})$} & \colhead{Project ID} & \colhead{RMS} \\
\cline{2-3}
\colhead{BGPS(ID)} & \colhead{$\alpha$(J2000)} & \colhead{$\delta$(J2000)} & \colhead{(\kms)} & \colhead{(kpc)} & \colhead{Refs\tablenotemark{a}} & & \colhead{(pc)} & \colhead{(K)} & \colhead{(\lsun)} & \colhead{(\msun)} & \colhead{(cm$^{-2}$)} &  & \colhead{(K)}
}
\colnumbers
\startdata
BG008.458-00.224(1363) & 18:05:22.61 & -21:44:43.5 & 38.0 & 4.30(0.37) & 1,2 & 2 & 0.52 & 14.5 & 2.9 & 2.8 & 22.6 & M22AP051 & 0.1 \\
BG009.212-00.202(1412) & 18:06:52.58 & -21:04:36.1 & 41.9 & 4.40(0.34) & 1,2 & 3 & 1.15 & 16.3 & 3.6 & 3.5 & 22.9 & M22AP051 & 0.13 \\
BG010.214-00.324(1466) & 18:09:24.75 & -20:15:36.2 & 9.4 & 4.8(0.2) & 3 & 2 & 2.02 & 25.5 & 5.0 & 3.8 & 22.9 & M22AP051 & 0.08 \\
BG010.416-00.030(1491) & 18:08:44.03 & -19:56:27.3 & 67.8 & 8.21(0.12) & 4 & 2 & 0.39 & 20.4 & 3.2 & 3.0 & 22.0 & M22AP051 & 0.13 \\
BG010.681-00.028(1518) & 18:09:16.34 & -19:42:29.2 & 50.8 & 3.85(0.26) & 4 & 2 & 0.62 & 20.1 & 3.5 & 2.9 & 22.7 & M22AP051 & 0.09 \\
BG011.083-00.536(1584) & 18:11:59.17 & -19:36:03.9 & 29.8 & 2.85(0.21) & 4,5 & 1 & 0.5 & 11.8 & 2.3 & 3.1 & 23.0 & M22AP051 & 0.1 \\
BG012.889+00.490(1803) & 18:11:51.27 & -17:31:27.3 & 32.1 & 2.34(0.12) & 6 & 3 & 0.84 & 23.4 & 4.2 & 3.0 & 23.3 & M16AP067 & 0.1 \\
BG013.816+00.003(1943) & 18:15:30.23 & -16:56:36.9 & 47.6 & 3.86(0.22) & 4 & 2 & 0.18 & 18.1 & 2.3 & 2.3 & 22.0 & M22AP051 & 0.08 \\
BG013.882-00.143(1956) & 18:16:10.32 & -16:57:18.3 & 18.1 & 2.23(0.45) & 7 & 2 & 0.24 & 17.8 & 2.6 & 2.2 & 23.5 & M22AP051 & 0.08 \\
BG014.606+00.012(2072) & 18:17:02.39 & -16:14:40.1 & 26.8 & 2.78(0.56) & 7 & 3 & 0.58 & 24.1 & 3.8 & 2.7 & 22.7 & M22AP051 & 0.13 \\
BG014.708-00.224(2097) & 18:18:06.42 & -16:15:59.9 & 37.4 & 3.40(0.48) & 8 & 1 & 0.46 & 9.7 & 1.7 & 3.0 & 22.7 & M22AP051 & 0.1 \\
BG015.021-00.620(2153) & 18:20:10.63 & -16:10:41.0 & 19.9 & 1.85(0.10) & 3 & 2 & 0.58 & 26.4 & 4.1 & 2.9 & 22.8 & M22AP051 & 0.12 \\
BG015.123-00.558(2175) & 18:20:08.99 & -16:03:31.9 & 18.7 & 1.85(0.10) & 3 & 1 & 0.66 & 24.7 & 3.6 & 2.5 & 22.2 & M22AP051 & 0.09 \\
BG016.894+00.486(2314) & 18:19:47.96 & -14:00:17.3 & 23.9 & 1.87(0.14) & 4 & 1 & 0.06 & 14.1 & 1.4 & 1.8 & 22.4 & M22AP051 & 0.09 \\
BG023.875+00.534(3184) & 18:32:53.71 & -07:48:26.4 & 95.8 & 5.25(0.66) & 4 & 2 & 0.59 & 12.1 & 2.6 & 3.3 & 22.4 & M22AP051 & 0.09 \\
BG023.968-00.110(3202) & 18:35:22.58 & -08:01:17.8 & 71.9 & 4.66(0.93) & 7 & 3 & 1.11 & 20.1 & 4.0 & 3.4 & 22.9 & M22AP051 & 0.08 \\
BG024.010+00.488(3210) & 18:33:18.69 & -07:42:31.6 & 94.1 & 5.10(0.61) & 4,5 & 0 & 0.74 & 10.3 & 2.1 & 3.5 & 22.9 & M22AP051 & 0.1 \\
BG024.329+00.142(3284) & 18:35:08.64 & -07:35:06.4 & 114.8 & 5.78(0.29) & 8 & 3 & 1.21 & 23.8 & 4.5 & 3.4 & 23.1 & M22AP051 & 0.09 \\
BG024.414+00.102(3313) & 18:35:26.71 & -07:31:41.1 & 113.2 & 7.91(0.89) & 4,7,8 & 1 & 2.45 & 17.7 & 4.3 & 4.0 & 22.8 & M22AP051 & 0.09 \\
BG025.400-00.141(3507) & 18:38:08.51 & -06:45:50.4 & 95.6 & 5.63(0.76) & 4 & 3 & 1.42 & 34.4 & 5.5 & 3.5 & 22.7 & M22AP051 & 0.18 \\
BG027.317+00.175(3777) & 18:40:32.76 & -04:54:55.5 & 33.5 & 12.38(0.49) & 4,5,8 & 2 & 1.38 & 19.9 & 4.1 & 3.6 & 22.4 & M22AP051 & 0.08 \\
BG028.341+00.140(3938) & 18:42:33.05 & -04:01:15.4 & 80.1 & 4.38(0.27) & 4,8 & 2 & 0.21 & 17.4 & 2.5 & 2.2 & 22.1 & M22AP051 & 0.09 \\
BG028.565-00.236(3998) & 18:44:18.14 & -03:59:36.8 & 86.5 & 4.68(0.38) & 4,8 & 1 & 1.91 & 17.0 & 2.3 & 3.0 & 23.8 & M22AP051 & 0.08 \\
BG029.397-00.095(4152) & 18:45:19.35 & -03:11:20.4 & 105.8 & 8.02(0.59) & 4,7 & 2 & 1.01 & 15.4 & 3.5 & 3.5 & 22.8 & M22AP051 & 0.08 \\
BG030.719-00.081(4518) & 18:47:41.32 & -02:00:22.8 & 93.2 & 5.20(0.58) & 4,5,7,8 & 3 & 1.16 & 21.6 & 4.7 & 3.8 & 23.3 & M22AP051 & 0.08 \\
BG030.772-00.801(4539) & 18:50:21.01 & -02:17:15.0 & 79.2 & 4.10(0.36) & 5,7,8 & 3 & 0.48 & 19.8 & 3.0 & 2.5 & 22.3 & M22AP051 & 0.08 \\
BG033.740-00.017(5252) & 18:52:58.34 & 00:42:41.1 & 105.5 & 6.08(0.58) & 4,5,8 & 1 & 1.8 & 13.9 & 3.5 & 3.7 & 23.0 & M22AP051 & 0.08 \\
BG034.259+00.222(5341) & 18:53:004.1 & 01:16:56.3 & 57.7 & 9.81(0.38) & 4 & 1 & 1.38 & 15.4 & 3.1 & 3.8 & 22.2 & M22AP051 & 0.1 \\
BG034.591+00.244(5414) & 18:53:35.75 & 01:35:16.1 & -24.1 & 15.03(0.49) & 4,5,8 & 3 & 0.71 & 23.0 & 4.1 & 3.2 & 22.1 & M22AP051 & 0.08 \\
BG034.712-00.596(5433) & 18:56:48.45 & 01:18:44.2 & 44.6 & 2.62(0.54) & 4,5,8 & 3 & 0.5 & 22.1 & 3.5 & 2.6 & 22.7 & M22AP051 & 0.09 \\
BG036.840-00.022(5798) & 18:58:39.19 & 03:28:03.1 & 58.3 & 9.54(0.45) & 4,7 & 3 & 1.2 & 16.9 & 3.7 & 3.5 & 22.6 & M22AP051 & 0.08 \\
BG039.267-00.589(5973) & 19:05:08.01 & 05:21:56.3 & 62.9 & 4.21(0.54) & 4,7 & 2 & 0.39 & 14.6 & 2.6 & 2.8 & 22.6 & M22AP051 & 0.09 \\
BG043.121+00.033(6116) & 19:10:03.61 & 09:04:24.1 & 7.6 & 11.24(0.43) & 4,7,8 & 2 & 3.43 & 22.1 & 5.2 & 4.1 & 22.4 & M22AP051 & 0.09 \\
BG044.661+00.351(6159) & 19:11:048.3 & 10:35:09.5 & 19.1 & 10.49(0.32) & 4 & 2 & 0.5 & 22.8 & 3.8 & 3.1 & 22.1 & M22AP051 & 0.11 \\
BG049.210-00.342(6313) & 19:23:01.23 & 14:17:04.2 & 66.6 & 5.37(0.41) & 4,7,8 & 2 & 1.64 & 28.0 & 5.3 & 3.7 & 22.7 & M22AP051 & 0.13 \\
BG081.721+00.572(6909) & 20:39:00.64 & 42:22:48.8 & -1.1 & 1.50(0.08) & 9 & 3 & 0.24 & 31.0 & 3.6 & 2.2 & 24.1 & M19BP033 & 0.08 \\
BG133.748+01.197(7364) & 02:25:53.13 & 62:04:08.2 & -38.8 & 2.20(0.72) & 10 & 3 & 0.25 & 29.4 & 3.8 & 2.1 & 23.7 & M19BP033 & 0.14 \\
BG133.949+01.063(7380) & 02:27:04.39 & 61:52:18.4 & -46.7 & 2.20(0.72) & 10 & 3 & 0.35 & 19.9 & 3.0 & 2.2 & 23.3 & M19BP033 & 0.13 \\
\enddata
\tablecomments{Observing field in (1) is formatted as ``name(ID)'' in BGPS version 1.0.1. Equatorial coordinates at J2000 epoch are listed in (2)--(3). Velocities at the local standard of rest retrieved from the $J=3-2$ transition of HCO$^+$ and N$_2$H$^+$ \citep{Schlingman2011FollowLine,Shirley2013FollowLine} is listed in (4). Distances and their references are listed in (5)--(6). The evolutionary type is listed in (7). The clump properties including radius, dust temperature, bolometric luminosity, mass, and peak column density derived from ATLASGAL and HiGAL catalog are listed in (8)--(12). Project ID and derived spectral RMS noise at a channel width of 0.2\,\kms~in the central field are in (13)--(14).
}
\tablenotetext{a}{Distance references. [1] \citet{He2015Infall}; [2] \citet{Dunham2011SFproperty}; [3] \citet{Yuan2017HMSC}; [4] this work; [5] \citet{EB2015Distance}; [6] \citet{Xu2011Trigonmetric}; [7] \citet{Wienen2015ATLASGALdistance}; [8] \citet{Svoboda2016property}; [9] \citet{Rygl2012DR21distance}; [10] \citet{Navarete2011W3}.} 
\tablenotetext{b}{Evolutionary type retrieved from \citet{Urquhart2018Property}. 0 = quiescent, 1 = protostellar, 2 = young stellar object, 3 = associated with massive star forming indicators, including radio bright \hii~regions and masers.}
\end{deluxetable*}
\end{longrotatetable}

\startlongtable
\begin{deluxetable*}{ccccccccc}
\tabletypesize{\small}
\tablewidth{2pt}
\linespread{1.1}
\tablecaption{Properties of HCN Cores \label{tab:HCNcores}}
\tablehead{
\colhead{HCN Core\tablenotemark{a}} & \colhead{Flag} & \colhead{Xoffset} & \colhead{Yoffset} & \colhead{$\theta_{\rm maj}$} & \colhead{$\theta_{\rm min}$} & \colhead{PA} & \colhead{$F_{\rm peak}$} & \colhead{$R_{\rm core}$} \\
 & & \colhead{(arcsec)} & \colhead{(arcsec)} & \colhead{(arcsec)} & \colhead{(arcsec)} & \colhead{(deg)} & \colhead{(K$\cdot$\kms)} & (pc) 
}
\colnumbers
\startdata
BG008.458-00.224C1 & 1 & 14.1 & -16.5 & 26.15 & 16.78 & 141.9 & 3.7 & 0.13 \\
BG009.212-00.202C1 & 1 & 4.7 & -12.4 & 30.95 & 22.24 & 12.4 & 4.6 & 0.2 \\
BG010.214-00.324C1 & 1 & 2.8 & -16.2 & 34.45 & 26.0 & 92.4 & 7.4 & 0.26 \\
BG010.681-00.028C1 & 1 & 1.3 & -15.9 & 21.58 & 15.35 & 61.1 & 6.1 & 0.08 \\
BG011.083-00.536C1 & 1 & -1.7 & -9.5 & 29.42 & 22.63 & 82.5 & 6.2 & 0.13 \\
BG012.889+00.490C1 & 1 & 5.1 & -1.7 & 33.61 & 25.9 & 90.4 & 27.6 & 0.13 \\
BG013.882-00.143C1 & 1 & 5.4 & -11.8 & 16.31 & 11.05 & 143.7 & 4.2 & -- \\
BG014.606+00.012C1 & 1 & -1.9 & -13.2 & 32.87 & 18.51 & 79.6 & 11.9 & 0.11 \\
BG015.021-00.620C1 & 1 & 8.8 & -11.9 & 36.58 & 31.46 & 13.4 & 5.7 & 0.12 \\
BG016.894+00.486C1 & 1 & 12.4 & -23.1 & 22.74 & 11.14 & 147.1 & 1.5 & -- \\
BG023.875+00.534C1 & 1 & 11.3 & -4.9 & 25.3 & 20.91 & 14.3 & 15.9 & 0.2 \\
BG023.968-00.110C1 & 1 & -0.6 & -17.7 & 26.79 & 21.38 & 136.3 & 11.3 & 0.19 \\
BG024.010+00.488C1 & 1 & 4.3 & 0.1 & 15.53 & 11.59 & 107.7 & 4.9 & -- \\
BG024.329+00.142C1 & 1 & 5.0 & -2.4 & 23.92 & 19.35 & 26.1 & 32.5 & 0.2 \\
BG024.414+00.102C1 & 2 & 14.2 & -5.2 & 23.88 & 12.67 & 117.9 & 3.9 & -- \\
BG024.414+00.102C2 & 2 & -12.2 & -8.6 & 17.05 & 12.9 & 127.3 & 3.3 & -- \\
BG025.400-00.141C1 & 1 & 8.8 & -12.9 & 34.45 & 26.65 & 143.7 & 18.3 & 0.32 \\
BG027.317+00.175C1 & 1 & 2.9 & -17.6 & 16.92 & 12.37 & 12.1 & 5.0 & -- \\
BG028.341+00.140C1 & 1 & -1.2 & -11.0 & 20.85 & 12.49 & 42.2 & 3.2 & -- \\
BG028.565-00.236C1 & 1 & 2.0 & -15.0 & 34.14 & 30.41 & 142.2 & 3.5 & 0.28 \\
BG029.397-00.095C1 & 1 & 1.2 & -9.0 & 30.07 & 16.58 & 59.7 & 8.6 & 0.26 \\
BG030.719-00.081C1 & 2 & 14.7 & -19.9 & 33.09 & 23.37 & 9.2 & 16.6 & 0.26 \\
BG030.719-00.081C2 & 2 & -15.7 & -20.1 & 28.33 & 19.9 & 24.6 & 18.1 & 0.2 \\
BG030.772-00.801C1 & 1 & 13.5 & -17.2 & 28.92 & 24.59 & 112.3 & 10.4 & 0.19 \\
BG033.740-00.017C1 & 1 & 21.1 & -12.2 & 37.5 & 15.79 & 139.3 & 8.4 & 0.2 \\
BG034.259+00.222C1 & 1 & 17.3 & 25.0 & 12.09 & 8.87 & 150.2 & 2.0 & -- \\
BG034.712-00.596C1 & 1 & 11.1 & -4.9 & 25.27 & 21.14 & 15.1 & 15.9 & 0.1 \\
BG036.840-00.022C1 & 1 & 15.2 & -15.3 & 13.04 & 11.61 & 54.2 & 4.8 & -- \\
BG039.267-00.589C1 & 1 & 3.5 & -14.4 & 17.35 & 13.12 & 98.1 & 4.7 & -- \\
BG044.661+00.351C1 & 1 & 11.1 & -16.9 & 13.34 & 10.55 & 51.9 & 3.8 & -- \\
BG049.210-00.342C1 & 1 & 6.8 & -8.1 & 41.93 & 27.75 & 48.2 & 7.1 & 0.35 \\
BG081.721+00.572C1 & 1 & 6.3 & -7.8 & 34.97 & 32.47 & 136.6 & 80.8 & 0.1 \\
BG133.748+01.197C1 & 1 & 11.8 & -9.4 & 28.04 & 18.09 & 128.6 & 17.0 & 0.08 \\
BG133.949+01.063C1 & 1 & 9.7 & -7.9 & 33.44 & 29.27 & 62.7 & 62.8 & 0.13 \\
\enddata
\tablecomments{Core name and flag are listed in (1)--(2). Offsets along the x and y axes in the equatorial coordinate are listed in (3)--(4). Fitted parameters including FWHM major axis, minor axis, position angle, and peak flux are listed in (5)--(8). The core size deconvolved with the beam is listed in (9).
}
\tablenotetext{a}{The name of core is in format of ``Field+CN'' where ``Field'' is the name of fields in Table\,\ref{tab:sample} and ``CN'' donates the core ID (C1, C2,...).} 
\tablenotetext{b}{Flag for core detection. 0 = no detection; 1 = one core; 2 = two cores.}
\tablenotetext{c}{Unresolved cores are shown with ``--''.}
\end{deluxetable*}

\startlongtable
\begin{deluxetable*}{ccccccccc}
\tabletypesize{\small}
\tablewidth{2pt}
\linespread{1.1}
\tablecaption{Parameters of HCN Emission Lines \label{tab:HCNpar}}
\tablehead{
\colhead{HCN Core} & \colhead{$T_{\rm b,HCN}^\mathrm{pk}$} & \colhead{$V_{\rm LSR,HCN}$} & \colhead{$\mathrm{d}V_{\rm HCN}$} & \colhead{$V_{\rm peak,HCN}$} & \colhead{$V_{\rm sys}$} & \colhead{$\mathrm{d}V_{\rm thin}$\tablenotemark{a}} & \colhead{$\delta V$} & \colhead{Flag\tablenotemark{b}} \\
 & \colhead{(K)} & \colhead{(\kms)} & \colhead{(\kms)} & \colhead{(\kms)} & \colhead{(\kms)} & \colhead{(\kms)} & &
}
\colnumbers
\startdata
BG008.458-00.224C1 & 0.29(0.01) & 37.84(0.09) & 5.55(0.2) & 37.88 & 38.0 & 5.1 & -0.024 & S \\
BG009.212-00.202C1 & 0.23(0.01) & 42.10(0.14) & 6.83(0.32) & 41.14 & 41.9 & 5.8 & -0.131 & BP \\
BG010.214-00.324C1 & 0.25(0.01) & 12.69(0.08) & 6.61(0.18) & 13.39 & 11.7 & 5.3 & 0.319 & RP \\
BG010.416-00.030C1 & -- & -- & -- & -- & 67.8 & 4.2 & -- & N \\
BG010.681-00.028C1 & 0.27(0.01) & 51.41(0.1) & 5.83(0.24) & 50.95 & 50.8 & 3.8 & 0.04 & S \\
BG011.083-00.536C1 & 0.09(0.0) & 31.03(0.35) & 13.32(0.83) & 30.04 & 29.8 & 4.4 & 0.054 & S \\
BG012.889+00.490C1 & 0.46(0.0) & 33.04(0.03) & 7.12(0.07) & 32.43 & 33.4 & 3.0 & -0.323 & BP \\
BG013.816+00.003C1 & -- & -- & -- & -- & 47.6 & 2.3 & -- & N \\
BG013.882-00.143C1 & 0.20(0.01) & 17.29(0.16) & 9.65(0.38) & 17.78 & 18.1 & 4.5 & -0.071 & S \\
BG014.606+00.012C1 & 0.33(0.01) & 25.45(0.12) & 10.45(0.29) & 23.93 & 26.8 & 5.0 & -0.574 & BP \\
BG014.708-00.224C1 & -- & -- & -- & -- & 37.4 & 2.6 & -- & N \\
BG015.021-00.620C1 & 0.42(0.01) & 19.02(0.06) & 5.57(0.14) & 19.03 & 19.9 & 4.2 & -0.207 & S \\
BG015.123-00.558C1 & 0.04(0.01) & 20.43(0.38) & 5.31(0.9) & 18.71 & 18.7 & 2.2 & 0.004 & S \\
BG016.894+00.486C1 & 0.08(0.01) & 24.14(0.34) & 6.41(0.81) & 24.4 & 23.9 & 2.7 & 0.185 & S \\
BG023.875+00.534C1 & 0.46(0.01) & 96.98(0.05) & 6.36(0.11) & 97.89 & 94.7 & 4.5 & 0.709 & RP \\
BG023.968-00.110C1 & 0.12(0.0) & 71.94(0.19) & 12.98(0.46) & 71.0 & 71.9 & 6.4 & -0.141 & BP \\
BG024.010+00.488C1 & 0.14(0.01) & 95.34(0.37) & 15.22(0.86) & 96.41 & 94.1 & 3.9 & 0.592 & RP \\
BG024.329+00.142C1 & 0.23(0.01) & 114.31(0.11) & 10.73(0.27) & 114.69 & 114.8 & 4.4 & -0.025 & S \\
BG024.414+00.102C1 & 0.15(0.01) & 113.48(0.22) & 8.48(0.53) & 113.29 & 113.2 & 4.5 & 0.02 & S \\
BG024.414+00.102C2 & 0.14(0.01) & 112.63(0.3) & 10.39(0.71) & 112.49 & 113.2 & 4.5 & -0.158 & S \\
BG025.400-00.141C1 & 0.66(0.01) & 94.87(0.06) & 7.01(0.15) & 93.67 & 95.6 & 2.8 & -0.689 & BP \\
BG027.317+00.175C1 & 0.15(0.01) & 33.13(0.24) & 11.68(0.58) & 31.62 & 33.5 & 5.6 & -0.336 & S \\
BG028.341+00.140C1 & 0.07(0.01) & 81.40(0.52) & 14.31(1.22) & 83.76 & 80.1 & 3.4 & 1.076 & RP \\
BG028.565-00.236C1 & 0.09(0.0) & 86.28(0.29) & 14.48(0.69) & 84.85 & 86.5 & 4.5 & -0.367 & BP \\
BG029.397-00.095C1 & 0.11(0.0) & 105.85(0.22) & 11.58(0.52) & 102.91 & 105.8 & 4.8 & -0.602 & BP \\
BG030.719-00.081C1 & 0.59(0.01) & 92.82(0.05) & 9.45(0.13) & 90.75 & 93.2 & 7.6 & -0.322 & BP \\
BG030.719-00.081C2 & 0.61(0.01) & 92.14(0.05) & 9.26(0.11) & 90.55 & 93.2 & 7.6 & -0.349 & BP \\
BG030.772-00.801C1 & 0.13(0.0) & 76.34(0.18) & 12.65(0.42) & 76.94 & 79.2 & 4.4 & -0.514 & S \\
BG033.740-00.017C1 & 0.20(0.01) & 104.51(0.11) & 8.48(0.25) & 103.57 & 105.5 & 4.3 & -0.449 & BP \\
BG034.259+00.222C1 & -- & -- & -- & -- & 57.7 & 4.3 & -- & N \\
BG034.591+00.244C1 & -- & -- & -- & -- & -23.9 & 2.9 & -- & N \\
BG034.712-00.596C1 & 0.46(0.01) & 43.61(0.05) & 6.36(0.11) & 44.53 & 44.6 & 3.0 & -0.023 & S \\
BG036.840-00.022C1 & 0.34(0.01) & 57.59(0.08) & 7.23(0.2) & 58.35 & 58.3 & 4.5 & 0.011 & S \\
BG039.267-00.589C1 & 0.18(0.01) & 61.08(0.16) & 7.86(0.38) & 59.79 & 62.9 & 4.5 & -0.691 & BP \\
BG043.121+00.033C1 & -- & -- & -- & -- & 7.6 & 3.9 & -- & N \\
BG044.661+00.351C1 & 0.22(0.01) & 18.46(0.17) & 7.89(0.4) & 19.82 & 19.1 & 3.7 & 0.195 & S \\
BG049.210-00.342C1 & 0.51(0.01) & 64.43(0.05) & 5.22(0.11) & 65.11 & 66.6 & 3.3 & -0.452 & S \\
BG081.721+00.572C1 & 2.45(0.03) & -6.77(0.04) & 7.86(0.1) & -7.75 & -4.5 & 5.1 & -0.637 & BP \\
BG133.748+01.197C1 & 0.49(0.01) & -41.45(0.06) & 5.87(0.15) & -41.78 & -39.0 & 4.0 & -0.695 & BP \\
BG133.949+01.063C1 & 1.70(0.01) & -50.46(0.03) & 6.77(0.06) & -51.28 & -48.4 & 4.9 & -0.588 & BP \\
\enddata
\tablecomments{Core name is listed in (1). Fitted HCN spectral parameters including peak brightness temperature, velocity at local standard of rest, FWHM line width, and velocity at peak value are listed in (2)--(5) respectively. Systematic velocity and FWHM line width derived from the optical thin lines \citep{Shirley2013FollowLine} are in (6)--(7). Asymmetric parameter calculated by Eq.\,\ref{eq:velocity_difference}. HCN line profile identification is listed in (9). ``BP'' = blue profile, ``RP'' = red profile, and ``S'' = single-peaked profile, ``N'' = non-detection.
} 
\tablenotetext{a}{Masers are checked and used to correct the velocity \citep{Xi2015maser,Xi2016maser}}
\tablenotetext{b}{S = single component; B = blue profile; R = red profile; N = no detection.}
\end{deluxetable*}

\begin{deluxetable*}{ccccccc}
\tabletypesize{\small}
\tablewidth{6pt}
\linespread{1.1}
\tablecaption{Line Profiles at Multi-$J$ Transition Lines \label{tab:multi-J}}
\tablehead{
\multirow{5}{*}{\makecell[c]{Source Names}} & \multicolumn{3}{c}{Multi-$J$ Transition Lines} & \multicolumn{3}{c}{$\tau_{\rm HCN (4-3)}$} \\
& \multicolumn{3}{c}{Critical density at 20\,K (cm$^{-3}$)} & \multicolumn{3}{c}{at collision partner density of} \\
\cmidrule(r){2-4} \cmidrule(r){5-7}
& \colhead{\hcopoz} & \colhead{\hcoptt} & \colhead{\hcnft} & \multicolumn{3}{c}{$n_{\rm H_2}$ (cm$^{-3}$)} \\
& \colhead{$4.5\times10^{4}$} & \colhead{$1.4\times10^6$} & \colhead{$2.3\times10^7$} & \colhead{$10^5$} & \colhead{$10^{5.5}$} & \colhead{$10^6$}
}
\colnumbers
\startdata
BG008.458-00.222 & BP & BP & S & 7.8(-3) & 2.5(-2) & 7.7(-2) \\
BG008.670-00.356 & BP & BP & -- & -- & -- & -- \\
BG009.212-00.202 & RP & BP & BP & 3.9(-1) & 1.4(0) & 3.2(0) \\
BG011.083-00.536 & BP & BP & S & 6.0(-3) & 1.8(-2) & 5.6(-2) \\
BG012.889+00.490 & RP & BP & BP & 5.6(0) & 1.1(1) & 1.3(1) \\
BG014.633-00.574 & BP & BP & -- & -- & -- & -- \\
\enddata
\tablecomments{Source names are inherited from Column (1) of Table\,\ref{tab:sample}. Profiles in multi-$J$ transition lines are shown in columns (2)--(4). The critical density at 20\,K is listed below the transition lines \citep{Shirley2015Lines}. ``BP'' = blue profile, ``RP'' = red profile, and ``S'' = single-peaked profile. Opacity of \hcnft~lines at three different volume densities of the collision partner H$_2$ is shown in columns (5)--(6). The value of opacity is shown in the form of ``a(b)'', donating $a\times10^b$.}
\end{deluxetable*}

\startlongtable
\begin{deluxetable*}{cccccccc}
\tabletypesize{\small}
\tablewidth{2pt}
\linespread{1.1}
\tablecaption{Hill5 Fitting Result of Blue-profile HCN (4-3) lines \label{tab:Hill5}}
\tablehead{
\colhead{HCN Core} & \colhead{Range} & \multicolumn{5}{c}{Hill5 Fitting Results} & \colhead{$\dot{M}$} \\
& \colhead{(\kms)} & \colhead{$\tau_{\rm core}$} & \colhead{$v_{\rm LSR}$ (\kms)} & \colhead{$v_{\rm infall}$ (\kms)} & \colhead{$\sigma$ (\kms)} & \colhead{$T_{\rm peak}$ (K)} & \colhead{($10^{-3}$\,\massrate)}
}
\colnumbers
\startdata
BG009.212-00.202C1 & [40,45] & 2.52(0.21) & 42.39(0.05) & 0.25(0.06) & 1.18(0.05) & 5.07(0.04) & 0.58(0.15) \\
BG012.889+00.490C1 & [29,37] & 1.55(0.12) & 33.46(0.07) & 1.20(0.15) & 1.78(0.04) & 5.95(0.03) & 4.47(0.72) \\
BG014.606+00.012C1 & [22,30] & 2.61(0.22) & 25.91(0.07) & 0.19(0.09) & 1.82(0.07) & 5.64(0.05) & 0.15(0.07) \\
BG025.400-00.141C1 & [92,98] & 1.52(0.16) & 94.98(0.06) & 0.61(0.13) & 1.51(0.06) & 6.92(0.04) & 1.41(0.32) \\
BG028.565-00.236C1 & [82,92] & 2.63(0.48) & 87.31(0.26) & 1.26(0.35) & 1.98(0.17) & 4.00(0.05) & 32.1(9.4)  \\
BG030.719-00.081C2 & [88,98] & 1.83(0.10) & 92.81(0.05) & 1.06(0.10) & 2.14(0.04) & 6.59(0.02) & 7.91(1.10) \\
BG033.740-00.017C1 & [100,112] & 1.73(1.35) & 105.58(0.12) & 1.59(1.75) & 1.75(0.63) & 4.70(0.20) & 4.58(5.04) \\
BG039.267-00.589C1 & [58,66] & 1.40(0.43) & 61.62(0.14) & 1.00(0.45) & 1.74(0.16) & 4.66(0.08) & $<$1.60(0.74) \\
BG081.721+00.572C1 & [-10,-2] & 2.45(0.09) & -6.04(0.04) & 1.19(0.07) & 1.57(0.03) & 11.78(0.07) & 21.5(2.46) \\
BG133.748+01.197C1 & [-44,-38] & 1.09(0.30) & -41.24(0.08) & 0.74(0.32) & 1.39(0.10) & 6.35(0.22) & 4.26(1.89) \\
BG133.949+01063C1 & [-54,-46] & 2.08(0.42) & -49.63(0.06) & 1.46(0.46) & 1.45(0.19) & 9.55(0.10) & 5.48(1.79) \\
\enddata
\tablecomments{Core name is listed in (1). Velocity range used for model fitting is listed in (2). Hill5 fitting results including optical depth, velocity at local standard of rest, infall velocity, velocity dispersion, and peak excitation temperature are listed in (3)--(7). The mass infall rate is listed in (8).
}
\end{deluxetable*}

\clearpage

\section*{Acknowledgment}
\begin{acknowledgements}

We thank the anonymous referee for the constructive comments. 

FWX and KW acknowledge support from the National Science Foundation of China (11721303, 11973013, 12033005),
the China Manned Space Project (CMS-CSST-2021-A09), 
National Key R\&D Program of China (2022YFA1603102), 
and the High-Performance Computing Platform of Peking University. 
YXH acknowledges support from the Chinese Academy of Sciences (CAS) ``Light of West China'' Program (2020-XBQNXZ-017). 

This research used the facilities of the Canadian Astronomy Data Centre operated by the National Research Council of Canada with the support of the Canadian Space Agency. The James Clerk Maxwell Telescope is operated by the East Asian Observatory on behalf of The National Astronomical Observatory of Japan; Academia Sinica Institute of Astronomy and Astrophysics; the Korea Astronomy and Space Science Institute; the National Astronomical Research Institute of Thailand; Center for Astronomical Mega-Science (as well as the National Key R\&D Program of China with No. 2017YFA0402700). Additional funding support is provided by the Science and Technology Facilities Council of the United Kingdom and participating universities and organizations in the United Kingdom and Canada. The data used in this paper are from project M16AP067, M19BP033, and M22AP051. We thank Junhao Liu for great help on the JCMT observations.

\textit{Software}. This research uses \astropy, a community-developed core Python package for Astronomy \citep{Astropy2013,Astropy2018,Astropy2022}. This research makes use of \montage, funded by the National Science Foundation under Grant Number ACI-1440620, and previously funded by the National Aeronautics and Space Administration's Earth Science Technology Office, Computation Technologies Project, under Cooperative Agreement Number NCC5-626 between NASA and the California Institute of Technology. This research has used the SIMBAD database, operated at CDS, Strasbourg, France \citep{2000A&AS..143....9W}. This research has used the program \SE, which builds a catalog of objects from an astronomical image \citep{1996A&AS..117..393B}. This research has used Python based package \psk~to fit spectral lines \citep{Ginsburg2011Pyspeckit,Ginsburg2022Pyspeckit}. This research has used \texttt{RADEX}, a computer program for fast non-LTE analysis of interstellar line spectra \citep{2007A&A...468..627V}.

\end{acknowledgements}

\clearpage
\appendix

\renewcommand\thefigure{\Alph{section}\arabic{figure}}    
\setcounter{figure}{0}
\section{The Definition of HCN Cores by \SE}
\label{app:sextractor}

The automatic source extraction program \SE~\citep{1996A&AS..117..393B} is used to extract sources from the observing fields. For each field, we first generate a RMS noise map from the line-free channels pixel-wise. The moment 0 (M0) maps, together with the corresponding RMS maps, serve as two inputs for \SE. 
Before the program runs, $\texttt{nthresh}=3$ is set to cut out the low SNR (\texttt{nthresh}$\times$local RMS) pixels. The deblending parameters (number of thresholds for deblending $\texttt{deblend\_nthresh}=512$, deblending contrast $\texttt{deblend\_cont}=10^{-5}$), and the parameter to control the minimum pixels in a core ($\texttt{min\_npix}=10$, that is, the number of pixels in a JCMT beam) are set for the source extraction procedure. Note that some of the fields, for example BG013.882-00.143, BG014.708-00.224, and BG015.021-00.620, have outlier tiles with abrupt background emission and RMS. However, such artifacts have no effect on source extraction, mostly due to a good background reduction and a local RMS map input of the \SE~program. When the program finishes, we discard the sources at the edge of fields, whose emission is not fully observed.

As a result, 32 fields have source detection while six have no detection. Among the 32 fields, two have detection of two sources. Since most sources are in shape of ellipses and centrally peaked, we call them HCN cores hereafter. 

All basic fitted parameters of HCN cores, including offsets (along the x and y axes) from the field center, FWHM along the major and minor axes ($\theta_{\rm maj}$ and $\theta_{\rm min}$), position angle (PA) and peak flux ($F_{\rm peak}$) are listed in column 3--8 of Table\,\ref{tab:HCNcores}. 

Shown in Figure\,\ref{fig:m0-1/2}, the HCN cores are marked with green ellipses whose major and minor axes are the FWHM of the source profile along the direction of the maximum and minimum dispersion, respectively. Green texts ``C1'' and ``C2'' indicate core ID(s) in one field. For those without detection of HCN cores, green dashed circles with diameter of 5 pixels \citep[$\sim30\arcsec$, i.e., the beamsize of SMT at 270\,GHz;][]{Shirley2013FollowLine} are used to outline the regions for extraction of spectra in Section\,\ref{result:avgspectra}. The non-detection fields have no fitted parameters, which are indicated by ``--'' in columns 3--8 of Table\,\ref{tab:HCNcores}.

\begin{figure*}[!ht]
\centering
\includegraphics[width=1.0\linewidth]{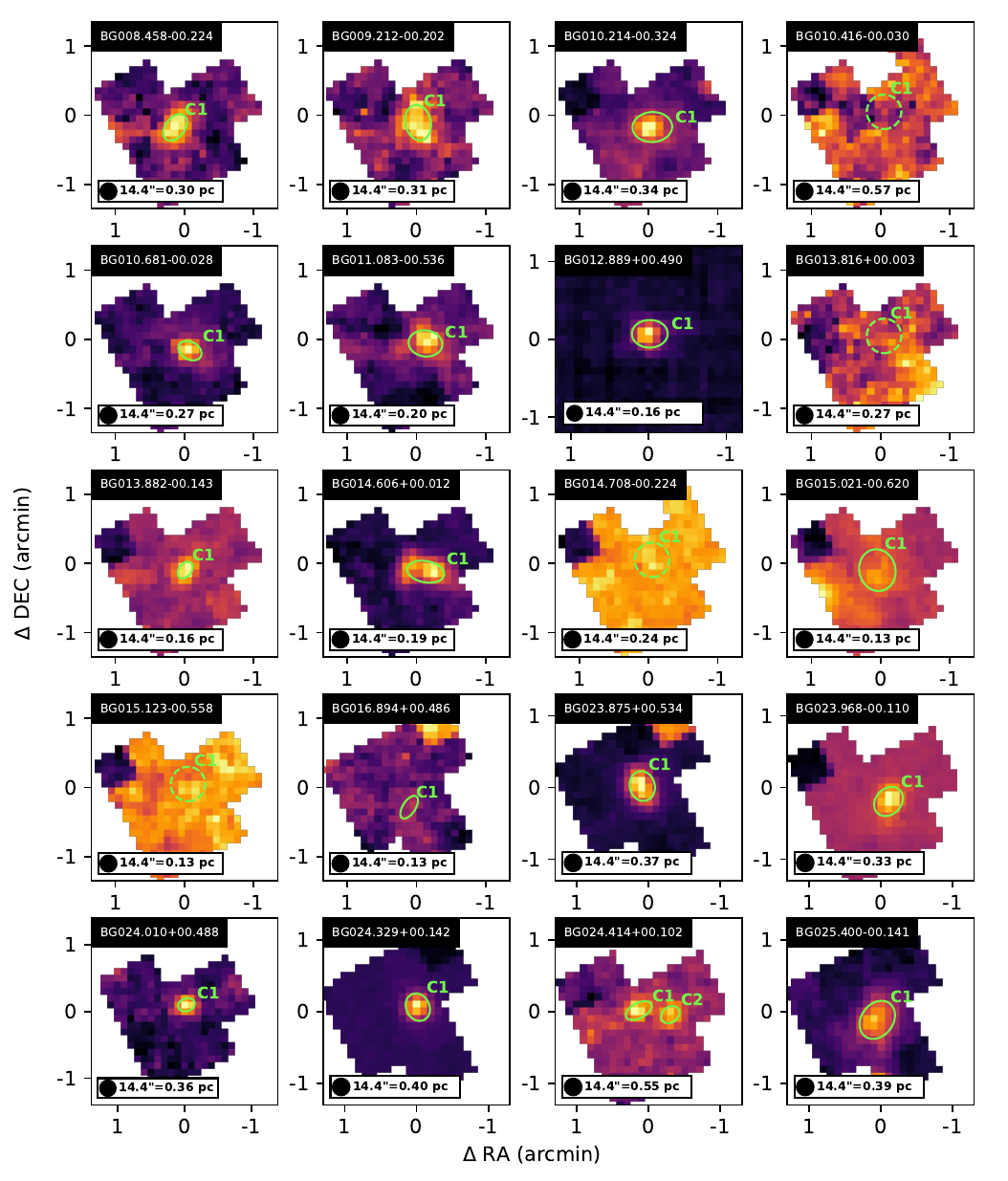}
\caption{The moment0 maps of \hcnft~in 38 fields. The names are labeled at the top left of each panel. The green solid ellipses mark the HCN cores by \SE, while the green dashed circles outline the regions where the spectra are extracted in the non-detection fields. The JCMT beam and its physical scale in each map are shown in the bottom left. \label{fig:m0-1/2}}
\end{figure*}
\addtocounter{figure}{-1}

\begin{figure*}[!ht]
\centering
\includegraphics[width=1.0\linewidth]{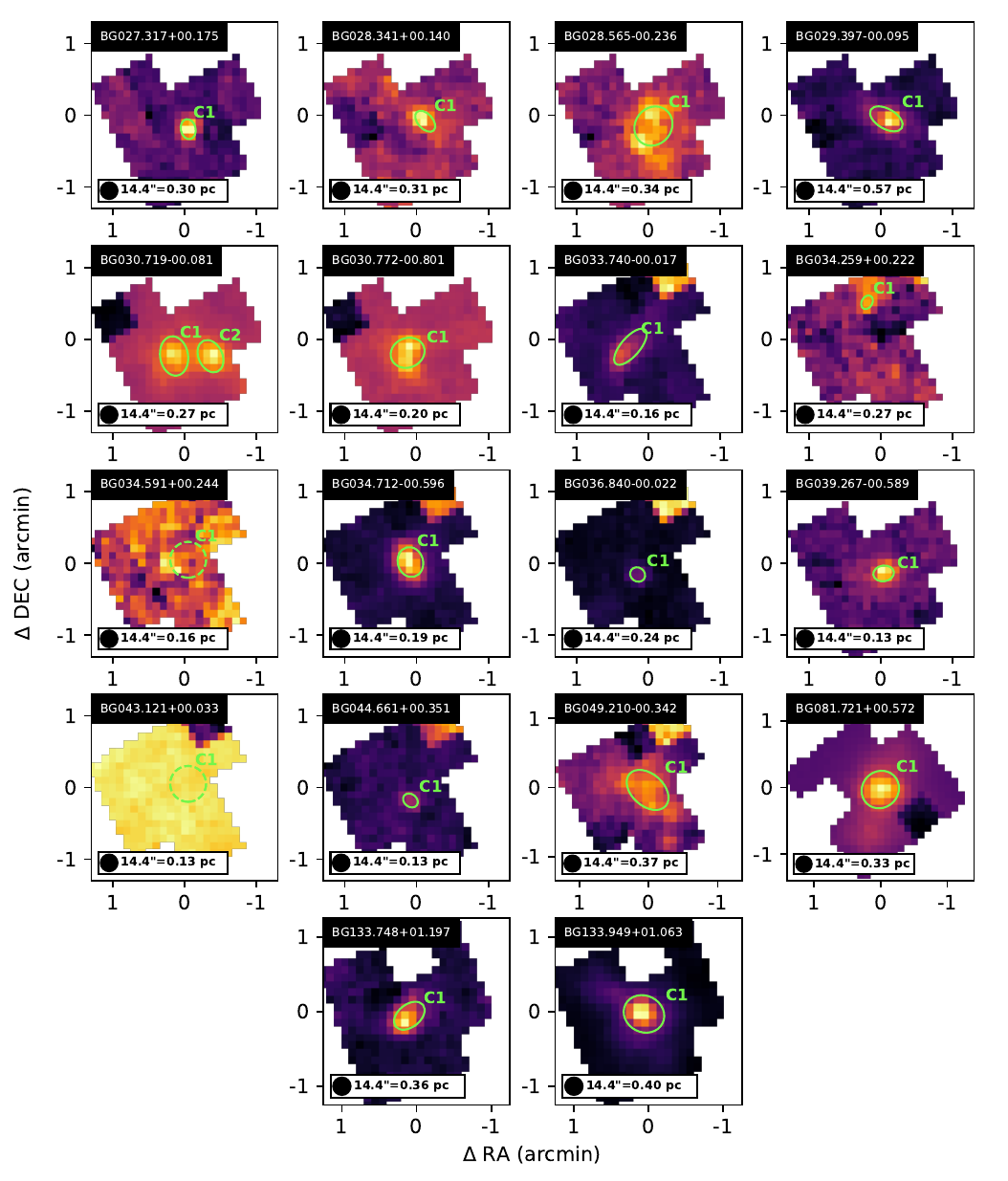}
\caption{Continued. \label{fig:m0-2/2}}
\end{figure*}

\clearpage

\renewcommand\thefigure{\Alph{section}\arabic{figure}}    
\setcounter{figure}{0}
\section{Core rotation identified by gradient of HCN (4-3)}
\label{app:rotate}

We present the spectral line maps of the 14 HCN cores exhibiting blue profiles in Figure\,\ref{fig:specgrid}. As deliberated in \citet{Redman2004Infall}, rotation would manifest blue-red asymmetries, leading to brighter blue peaks on one side of the core's rotation axis and brighter red peaks on the opposing side. Consequently, a pure rotational movement would yield a consistent velocity gradient perpendicular to the rotation axis. We compute the gradient of \hcnft~peak velocity within the core by employing the following formula,
\begin{equation}
\begin{aligned}
\Delta x_{\rm ij} &= \left(\frac{\partial V_{\rm peak,HCN}(x,y)}{\partial x}\right)_{\rm ij}\\
\Delta y_{\rm ij} &= \left(\frac{\partial V_{\rm peak,HCN}(x,y)}{\partial y}\right)_{\rm ij},
\end{aligned}
\end{equation}
where (i,j) indicates pixel location. The position angle of local maximum of gradient can be further calculated by,
\begin{equation}
\theta_{\rm ij} = \arctan\left(\frac{\Delta y_{\rm ij}}{\Delta x_{\rm ij}}\right).
\end{equation}
For an ideal core rotation model, the rotation axis should be perpendicular to the velocity gradient $\theta_{i,j}$. We identify three candidates of core rotation, BG023.968-00.110C1, BG029.397-00.095C1, and BG030.719-00.081C1, whose rotation axes have the position angles of 79\parcdeg4, 113\parcdeg8, and 101\parcdeg7. The rotation axes are labeled with green dashed lines in the panels of the three cores in Figure\,\ref{fig:specgrid}. 

\begin{figure*}[!h]
\centering
\includegraphics[width=0.32\linewidth]{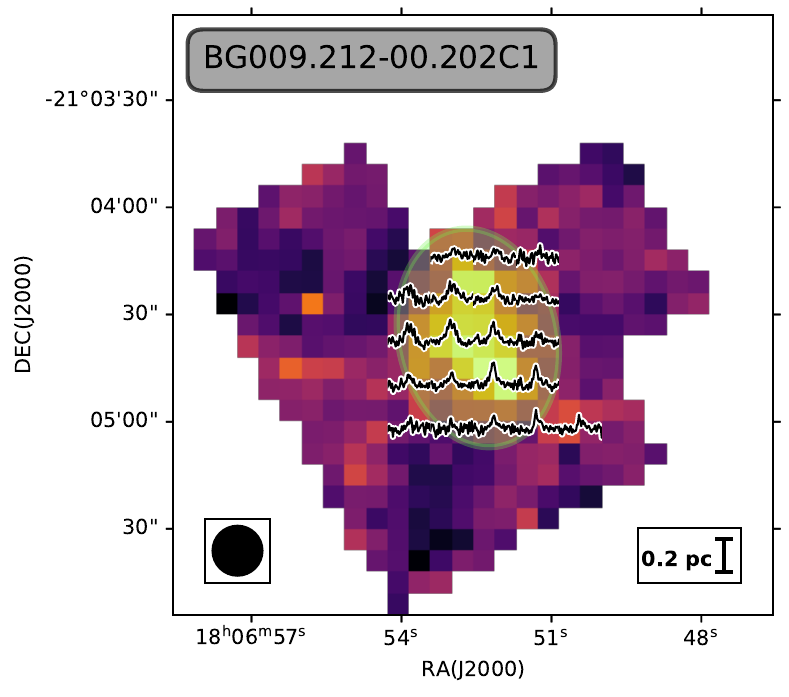}
\includegraphics[width=0.32\linewidth]{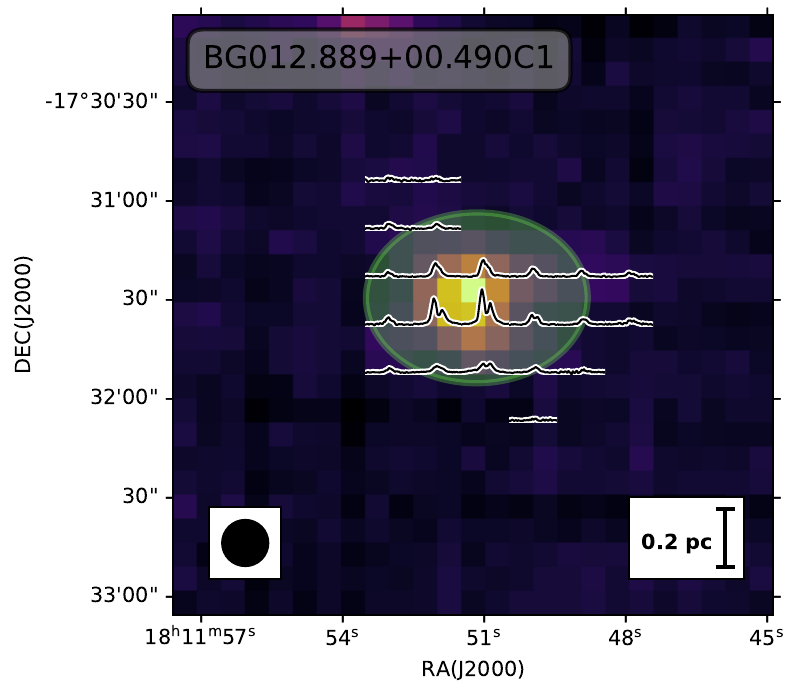}
\includegraphics[width=0.32\linewidth]{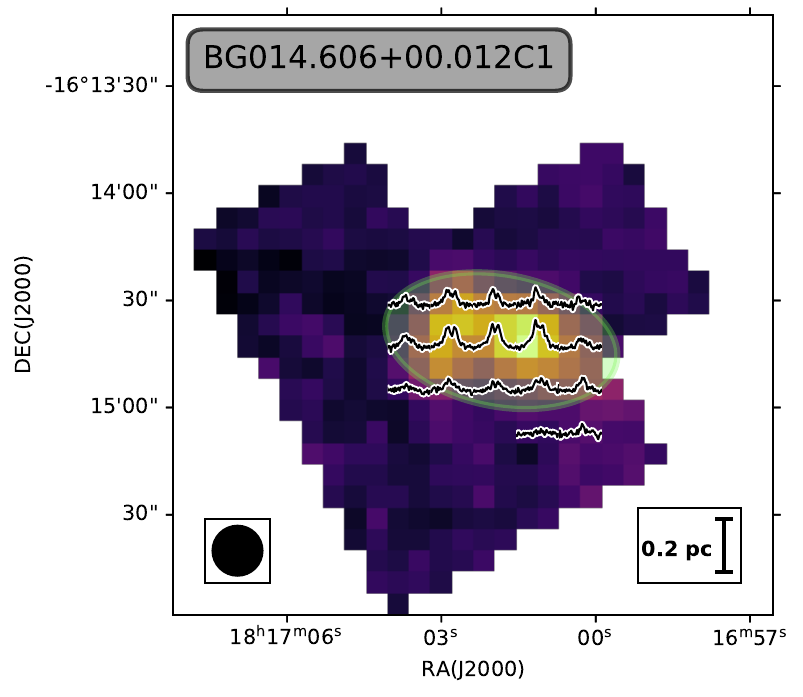}
\includegraphics[width=0.32\linewidth]{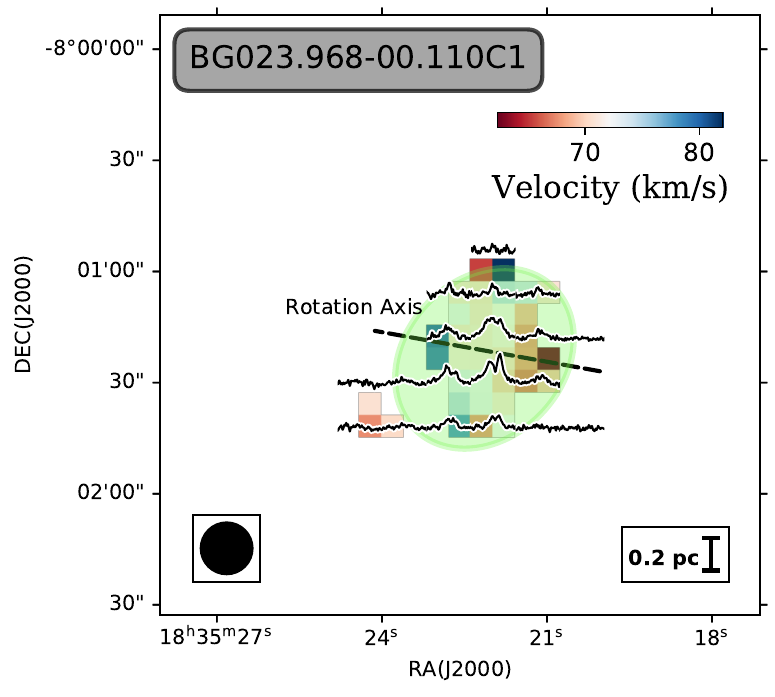}
\includegraphics[width=0.32\linewidth]{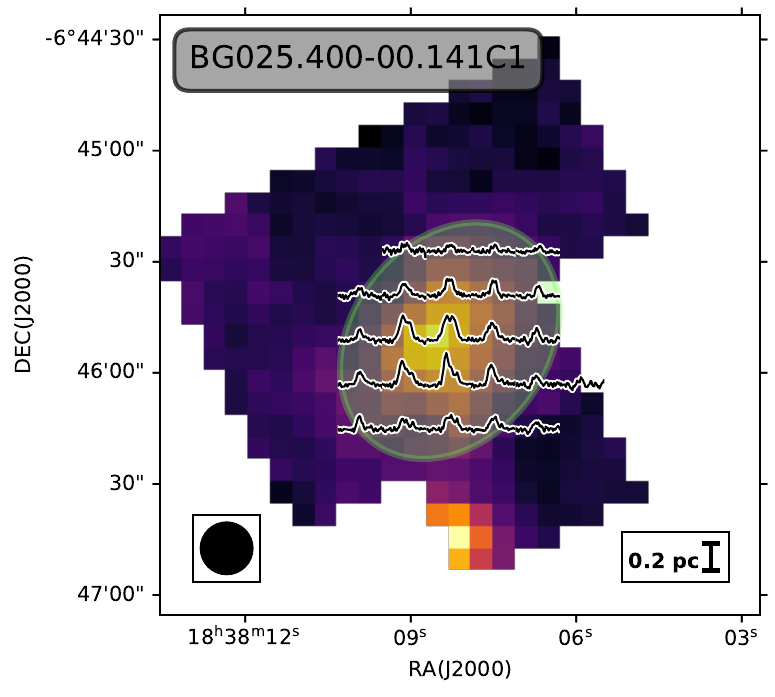}
\includegraphics[width=0.32\linewidth]{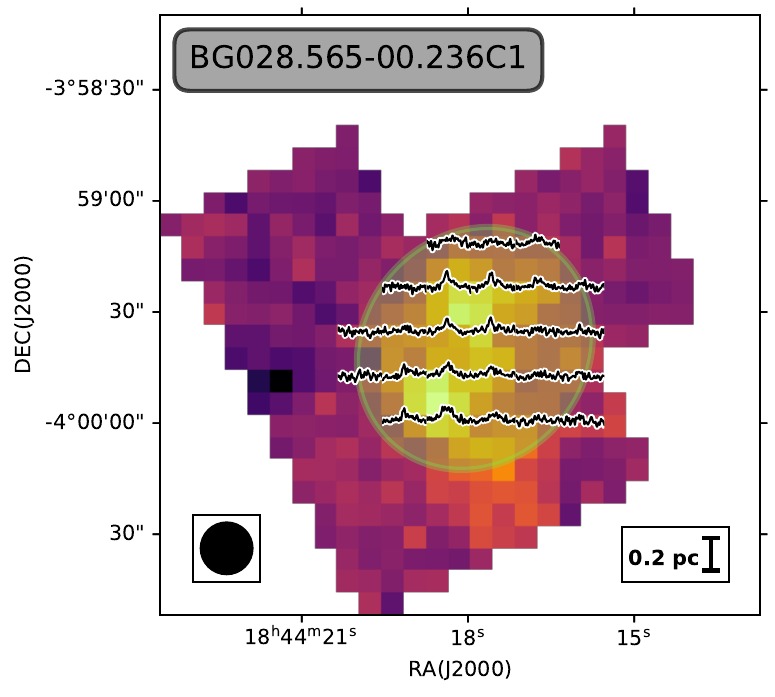}
\caption{For the 11 infall candidates, the line grids are overlaid on M0 color maps. Green ellipses mark the footprints of HCN cores. The lines shown are averaged from the $2\times2$ pix$^2$ box and smoothed to 0.4\,\kms. For the three core rotation candidates, black dashed lines show the rotation axes, which are overlaid on moment1 color maps. The JCMT beam and the scale bar of 0.2\,pc are shown on the bottom left and right, respectively. \label{fig:specgrid}}
\end{figure*}
\addtocounter{figure}{-1}

\begin{figure*}
\centering
\includegraphics[width=0.32\linewidth]{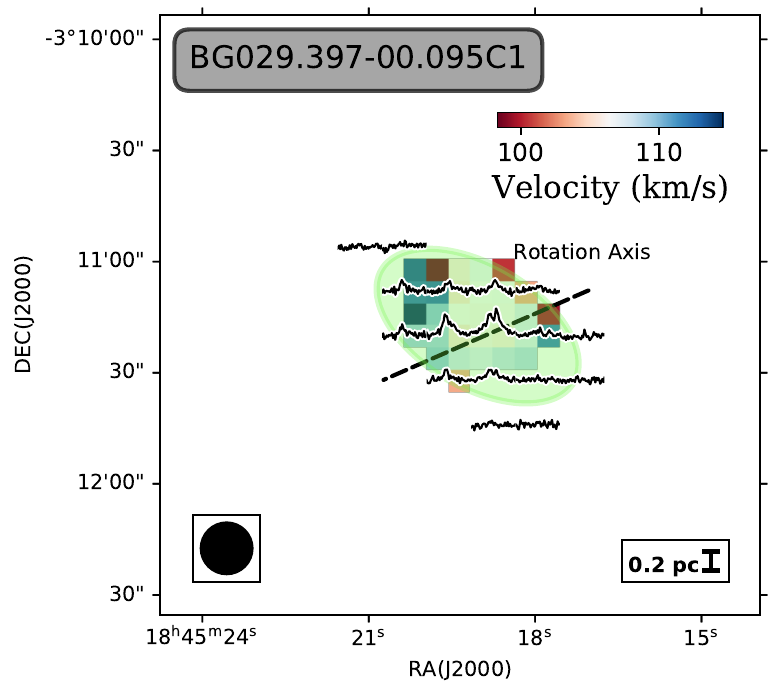}
\includegraphics[width=0.32\linewidth]{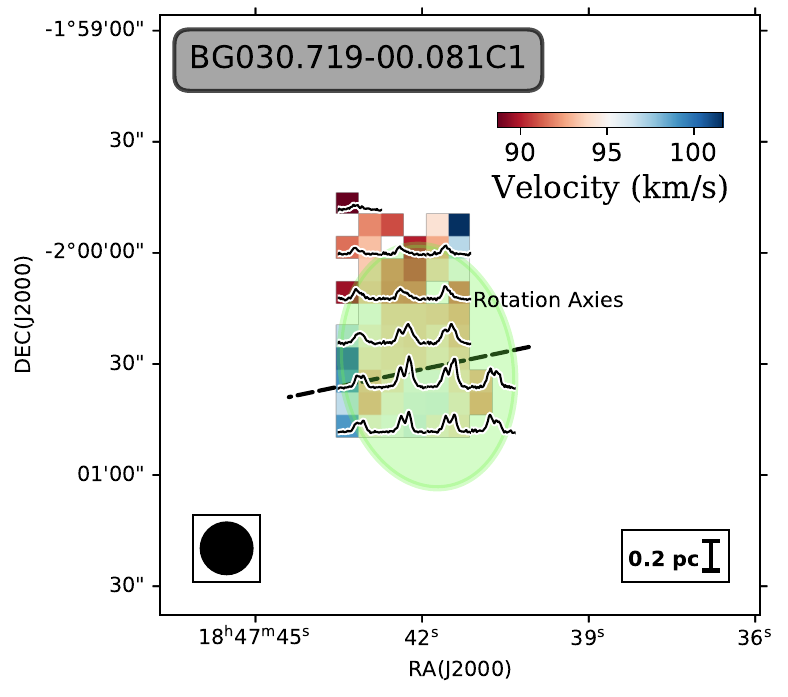}
\includegraphics[width=0.32\linewidth]{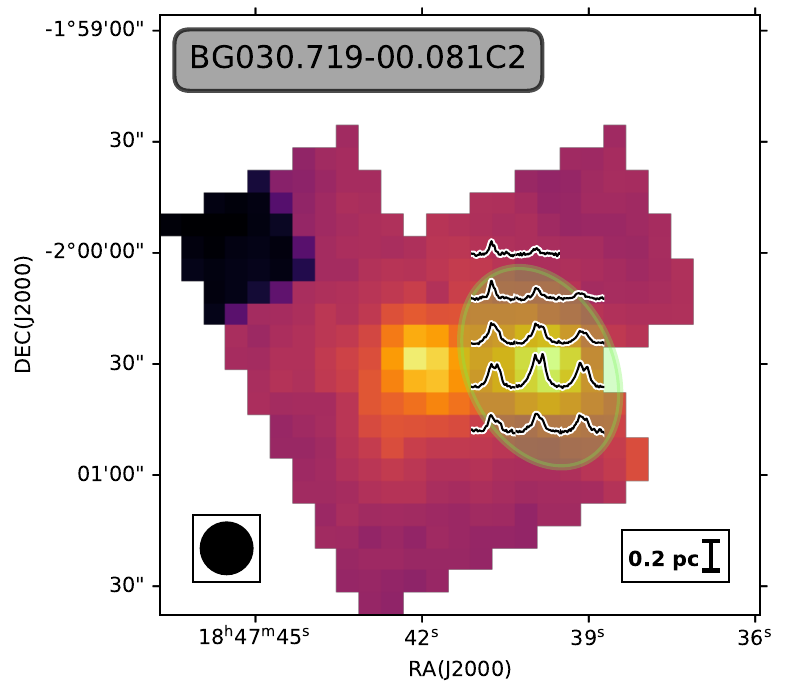}
\includegraphics[width=0.32\linewidth]{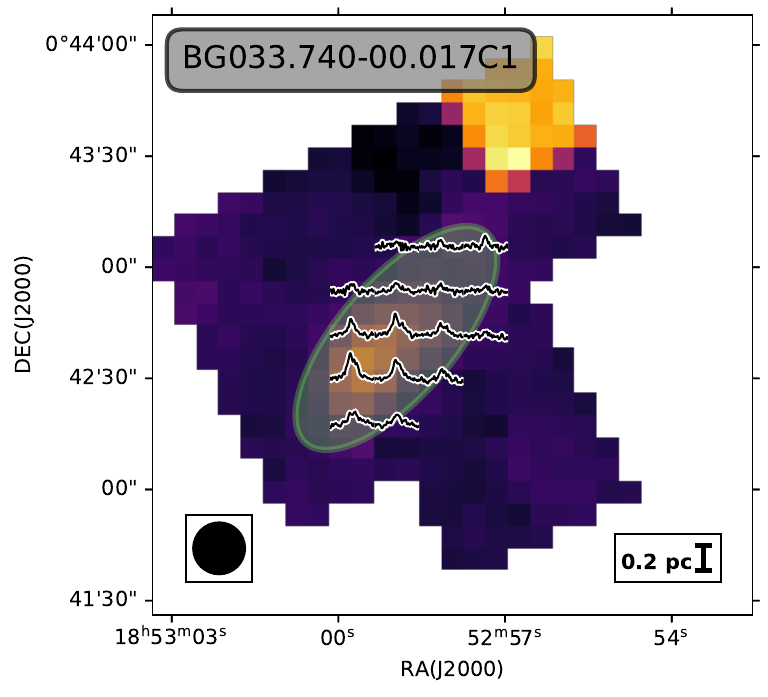}
\includegraphics[width=0.32\linewidth]{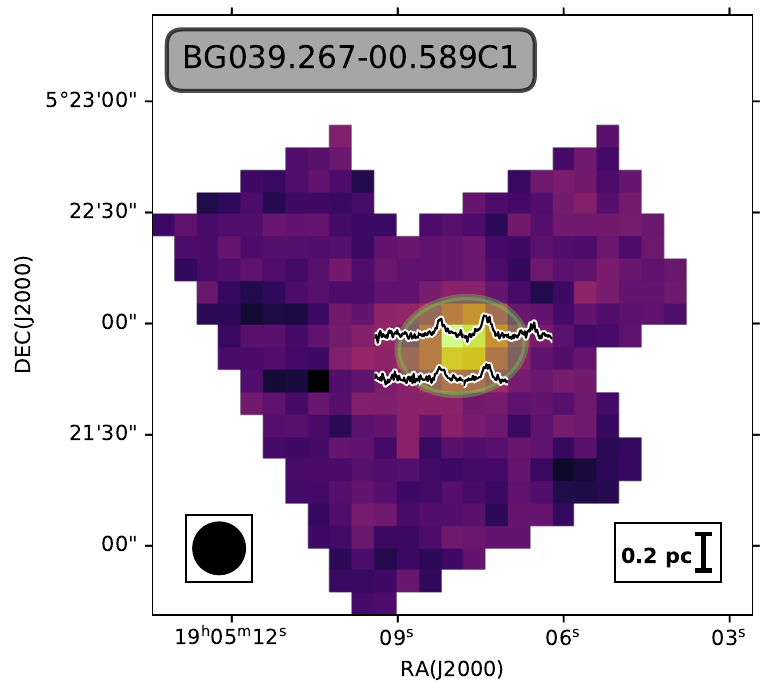}
\includegraphics[width=0.32\linewidth]{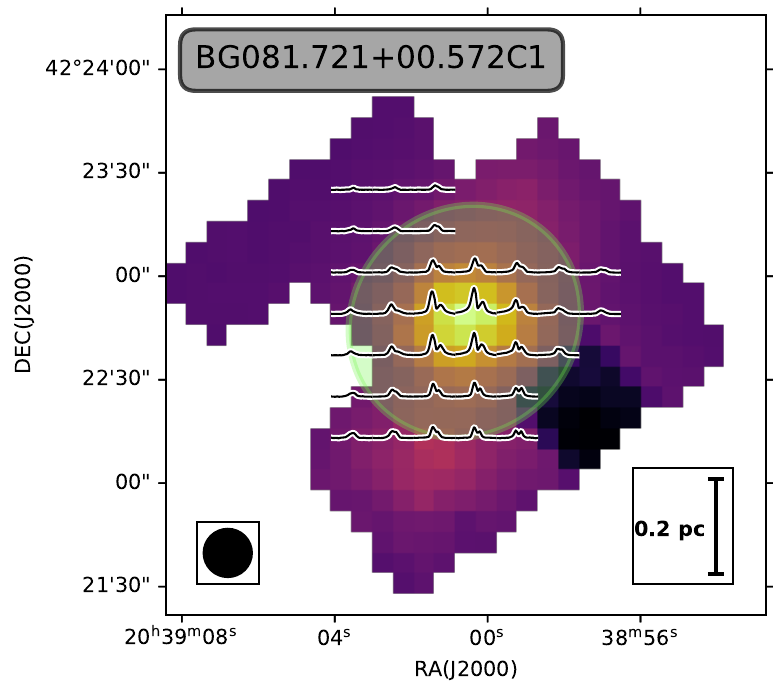}
\includegraphics[width=0.32\linewidth]{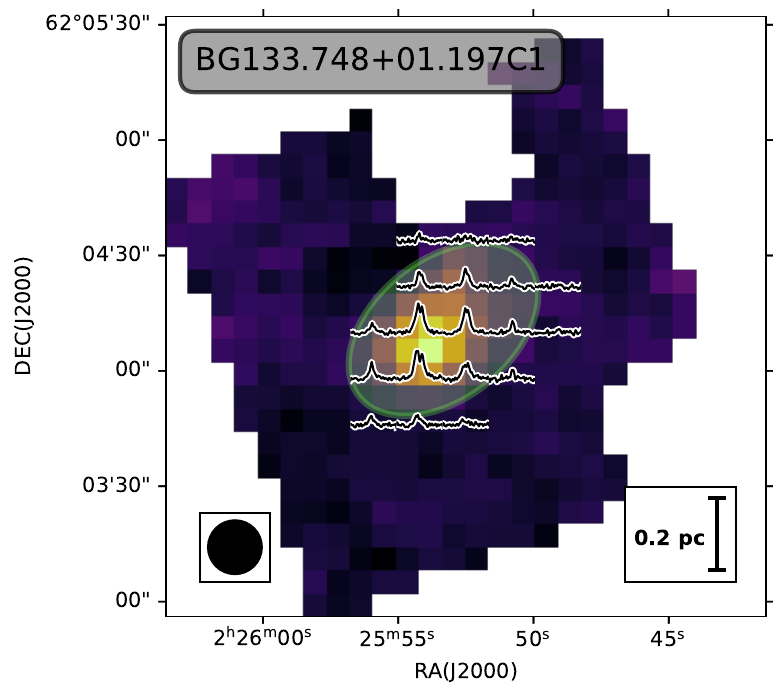}
\includegraphics[width=0.32\linewidth]{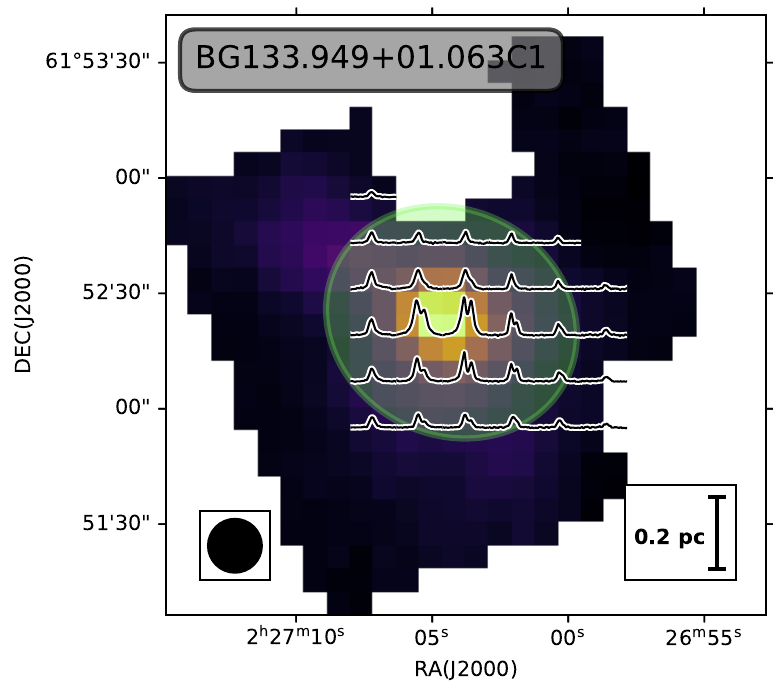}
\caption{Continued.}
\end{figure*}
\addtocounter{figure}{-1}
\clearpage

\section{Estimation of HCN Core Volume Density}
\label{app:vden}

A statistical study of 27 IRDCs by \citet{Peretto2023Decouple} found that the self-gravitating massive star-forming clumps have dynamically decoupled from their surrounding molecular clouds below the parsec scale, exhibiting a steeper density profile $\rho \propto r^{-2}$.
Adopting the density profile, we can derive the enclosed mass within a given radius $R$:
\begin{equation} \label{eq:M-R}
    M_{\rm enc} (<r) = \int^{r}_0 4\pi r^2 \rho \mathrm{d}r \propto r,
\end{equation}
in agreement with the IR-quiet protostellar MDCs found in Cygnus X and their hosted high-mass Class 0–like protostars \citep{Motte2007CygnusX,Bontemps2010CygnusX,Motte2018Review}. Therefore, the mean volume density of the enclosed mass inside radius $r$ writes,
\begin{equation} \label{eq:vden}
    n_{\rm H_2} (<r) = \frac{M_{\rm enc}(<r)}{(4/3)\pi\mu m_{\rm H} r^3} \propto \frac{1}{r^2},
\end{equation}
where $\mu = 2.81$ is the molecular weight per hydrogen molecule \citep{Evans2022SlowSF} and $m_{\rm H}$ is the mass of a hydrogen atom. Using Eq.\,\ref{eq:vden}, we can scale the volume density from the clump ($R_{\rm cl}$) down to the HCN cores ($R_{\rm core}$) as follows,
\begin{equation}
    n_{\rm H_2}(<R_{\rm core}) = \frac{R_{\rm cl}^2}{R_{\rm core}^2} n_{\rm H_2}(<R_{\rm cl}),
\end{equation}
where $n_{\rm H_2}(<R_{\rm cl}$ can be directly derived from clump mass $M_{\rm cl}$ and radius $R_{\rm cl}$ in columns (11) and (8), respectively.

\begin{figure}[!h]
\centering
\includegraphics[width=0.48\linewidth]{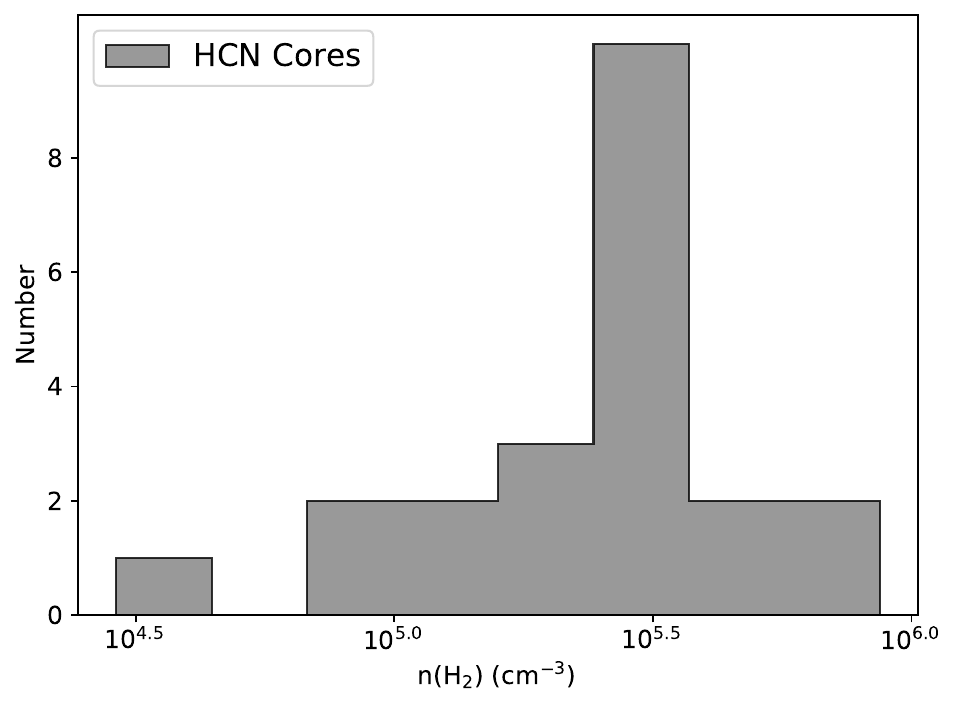}
\caption{Distribution of HCN core volume density. 86.4\% of HCN cores have a volume density within the range of $10^5$ to $10^6$\,cm$^{-3}$. \label{fig:vden}}
\end{figure}

Figure\,\ref{fig:vden} shows the distribution of HCN core volume density. The majority ($\sim90$\%) of HCN cores have a volume density within the range of $10^5$ to $10^6$\,cm$^{-3}$, which sets the input of the \texttt{RADEX} mock grid in Section\,\ref{multi-J:tau} and \ref{multi-J:triple}.

\clearpage

\section{The ``Hill5'' Fitting Model}
\label{app:hill5}

The ``Hill5'' fitting model assumes a core with a peak excitation temperature $T_{\rm peak}$ at the center and an excitation temperature of $T_0=T_{bg}=2.73$\,K at the proximal and distal of the core. The Planck temperature $J(T)$ is defined as,
\begin{equation}
J(T) = \frac{h\nu}{k_{\rm B}}\frac{1}{\exp(h\nu/k_{\rm B}T)-1}.
\end{equation}
We assume the $J(T)$ should drop linearly from $J(T_{\rm peak})$ at the center to $J(T_{\rm bg})$ at edges of the core, forming a hill in the $J(T)$ profile. 

In the unit of brightness temperature, the equation of radiation transfer writes as,
\begin{equation}
T_B = T_i e^{-\tau_0} + \int^{\tau_0}_0 J(T)e^{-\tau}\mathrm{d}\tau,
\end{equation}
where $T_i\equiv (c^2/2\nu^2k_{\rm B}) I_{\nu,i}$ is the incident specific intensity of radiation in unit of brightness temperature and $\tau_0$ is optical depth. 
Assuming a simple linear function $J(T)=J_1+[(J_2-J_1)/\tau_0]$, we can integrate the equation of radiation transfer to obtain,
\begin{equation}\label{eq:linear_solution}
T_B = T_i e^{-\tau_0} + (J_2-J_1)\frac{1-e^{-\tau_0}}{\tau_0} + J_1 - J_2 e^{-\tau_0}.
\end{equation}

In order to solve the equation of radiative transfer, we separate the core into two parts along the line of sight:
(1) the front part of the core in which the excitation temperature rises along the line of sight, whose optical depth is $\tau_{f}$; and (2) the rear part of the core in which the excitation temperature falls along the line of sight, whose optical depth is $\tau_{r}$. If both two parts infall with a velocity of $v_{\rm infall}$ relative to the systematic velocity $v_{\rm LSR}$, then the optical depth $\tau_{f}$ and $\tau_{r}$ as a function of line-of-sight velocity is written as,
\begin{equation}\label{eq:tau}
\begin{aligned}
\tau_{f}(v) &= \tau_{\rm core} \exp\left[-(v-v_{\rm LSR}-v_{\rm infall})^2/2\sigma^2\right],\\
\tau_{r}(v) &= \tau_{\rm core} \exp\left[-(v-v_{\rm LSR}+v_{\rm infall})^2/2\sigma^2\right],
\end{aligned}
\end{equation}
where $\tau_{\rm core}$ is the optical depth of the core and $\sigma$ is the velocity dispersion of each part of the core.

Substituting the first row of Eq.\,\ref{eq:tau} into Eq.\,\ref{eq:linear_solution}, the outgoing brightness temperature from the rear part writes,
\begin{equation}
T_{B,r} = J(T_{\rm bg}) e^{-\tau_r(v)} + \left[J(T_{\rm bg})-J(T_{\rm peak})\right]\frac{1-e^{-\tau_r(v)}}{\tau_r(v)} + J(T_{\rm peak}) - J(T_{\rm bg})e^{-\tau_r(v)},
\end{equation}
which then serves as the incident brightness temperature. Then the outgoing brightness temperature from the front part writes,
\begin{equation}
\Delta T_{B,f} = \left[J(T_{\rm peak})-J(T_{\rm bg})\right]\times\left[\frac{1-e^{-\tau_f(v)}}{\tau_f(v)}-\frac{(1-e^{-\tau_r(v)})}{\tau_r(v)}e^{-\tau_f(v)}\right]
\end{equation}
where the emission of reference point has been eliminated. Above all, the ``hill'' model contains five free parameters $\tau_{\rm core}$, $\sigma$, $T_{\rm peak}$, $v_{\rm LSR}$, and $v_{\rm infall}$ in total, which is the reason why it's called the ``Hill5'' model.

\clearpage
\bibliographystyle{aasjournal}

      \bibliography{JCMTINFALL}

\end{document}